\documentstyle[emulateapj]{article}
\input psfig.sty

\lefthead{Alonso-Herrero et al.}
\righthead{NGC~1614: A Laboratory for Starburst Evolution}
\begin{document}
\title{NGC~1614: A Laboratory for Starburst Evolution
\footnote{Based on observations with the NASA/ESA Hubble
Space Telescope, obtained at the Space Telescope Science Institute, which
is operated by the Association of Universities for Research in Astronomy,
Inc. under NASA contract No. NAS5-26555.}}

\author{A. Alonso-Herrero} 
\affil{Steward Observatory, The University of Arizona, Tucson, AZ 85721}
\affil{Present address: University of Hertfordshire, Department 
of Physical Sciences, College Lane, Hatfield, Herts AL10 9AB, UK}
\and
\author{C. W. Engelbracht, M. J. Rieke, G. H. Rieke, and A. C. Quillen}
\affil{Steward Observatory, The University of Arizona, Tucson, AZ 85721}

\begin{abstract}
The modest extinction and reasonably face-on viewing geometry make
the luminous infrared galaxy NGC~1614
an ideal laboratory for study of a powerful starburst.
{\it HST}/NICMOS observations show: 1.) deep  CO stellar absorption,
tracing a starburst nucleus about 45\,pc in diameter; 2.) surrounded
by a $\sim 600\,$pc diameter ring of supergiant H\,{\sc ii} regions revealed
in  Pa$\alpha$ line emission; 3.) lying within
a molecular ring indicated by its extinction
shadow in  $H-K$; 4.) all at the center of a disturbed
spiral galaxy.  The luminosities of the giant H\,{\sc ii} regions in the
ring are extremely high, an order of magnitude brighter than 30 Doradus;
very luminous H\,{\sc ii} regions, comparable with 30 Dor, are
also found in the spiral arms of the galaxy.
Luminous stellar clusters surround the nucleus and lie in the
spiral arms, similar to clusters observed in other infrared luminous
and ultraluminous galaxies. The star forming activity may have
been initiated by a merger between a disk galaxy
and a companion satellite, whose nucleus appears in
projection about 300\,pc to the NE of the nucleus
of the primary galaxy. The relation of deep stellar CO
bands to surrounding ionized gas ring to molecular gas indicates that
the luminous starburst started in the nucleus and is propagating
outward into the surrounding molecular ring. This hypothesis
is supported by evolutionary starburst modeling that shows
that the properties of NGC~1614 can be fitted with two
short-lived bursts of star formation separated by 5\,Myr (and
by inference by a variety of models with a similar duration
of star formation). The total dynamical mass of the starburst region
of $1.3 \times 10^9\,{\rm M}_\odot$ is mostly accounted for by the old pre-starburst
stellar population. Although our starburst models use
a modified Salpeter initial mass function (turning over near
1 M$_\odot$), the tight mass budget suggests that
the IMF may contain relatively more 10 - 30 M$_\odot$ stars
and fewer low mass stars than the Salpeter function.
The dynamical mass is nearly 4 times smaller than
the mass of molecular gas estimated from the standard ratio
of $^{12}$CO (1 $-$ 0) to H$_2$. A number of arguments
place the mass of gas in the starburst region at
$\sim 25$\% of the dynamical mass, nominally 
about 1/15 and with an upper limit of 1/10 
of the amount estimated from $^{12}$CO and
the standard ratio.


\end{abstract}


\section{INTRODUCTION}
Numerical simulations of collisions between gas-rich galaxies (e.g.,
Barnes \& Hernquist 1996 and references therein)
and even minor mergers between a gas-rich galaxy and a
satellite companion (e.g., Mihos \& Hernquist 1994)
show that these processes are
very efficient in transporting large quantities of molecular gas into the
galaxy nuclei. As a  result of the concentration of gas, a strong
burst (or various bursts) of star-formation  may occur, and
an AGN may be activated, as observed in many
infrared luminous and ultraluminous
galaxies (LIRGs and ULIRGs). Sanders et al. (1988) suggested that the
LIRG and ULIRG galaxies are  the
initial stage for the appearance of a quasar (see also the
review by Sanders \& Mirabel  1996). However, recent
results from mid-infrared spectroscopy seem to indicate that
most LIRGs and many ULIRGs may be powered by star formation
(Lutz et al. 1999). These objects therefore allow probing
the process of star formation on an extreme scale and intensity.

\begin{deluxetable}{lcccc}
\tablewidth{10cm}
\tablefontsize{\small}
\tablecaption{Log of the {\it HST} observations.}
\tablehead{\colhead{Camera} &  \colhead{Filter} & \colhead{scale} &
\colhead{$t_{\rm exp}$} & \colhead{filter/line}}
\startdata
NIC1   & F110M & 0.043 & 384 & pseudo-$J$\\
NIC2   & F160W & 0.076 & 192 & $H$ \\
NIC2   & F222M & 0.076 &640 & $K$ \\
NIC2   & F187N & 0.076 &640 & continuum \\
NIC2   & F190N & 0.076 & 640 & Pa$\alpha$ \\
NIC2   & F212N & 0.076 & 1792 & continuum \\
NIC2   & F215N & 0.076 & 1792 &  H$_2$ \\
NIC2   & F237M & 0.076 & 960 & CO absorption \\
WFPC2  & F606W & 0.041 &500 & $R$ \\
\enddata
\tablecomments{Column~(1) is the camera. Column~(2) is the filter. Column~(3)
the plate scale in arcsec pixel$^{-1}$. Column~(4) is the total
integration time in seconds. Column~(5) is the
corresponding ground-based filter, or the emission line.}
\end{deluxetable}

NGC~1614 (Arp~186) is relatively nearby
(distance $D = 64\,$Mpc for $H_0=75\,$km s$^{-1}$
Mpc$^{-1}$) and has a high infrared luminosity  ($L_{\rm IR} = 3 \times
10^{11}\,$L$_\odot$, which places this system in the
luminous infrared galaxy category). The galaxy
shows a spectacular outer structure with tidal
tails or plumes (see e.g., Neff et al. 1990 and also Figure~1)
suggesting that this morphology is  the result of an earlier
interaction/merger process with another galaxy. Neff et al.
(1990) collected optical, near-infrared and radio observations for this
galaxy and found no evidence for the presence of
active galactic nucleus (AGN), making it an excellent
laboratory for study of a very luminous starburst.

We present {\it HST}/NICMOS near infrared observations of NGC~1614,
giving unprecedented angular resolution on the galaxy
for this spectral region. These new observations
are combined with an archival {\it HST}/WFPC2 red image, 
ground-based near infrared spectroscopy, and results from the
literature to probe the starburst.

\begin{figure*}
\figurenum{1}
\plotfiddle{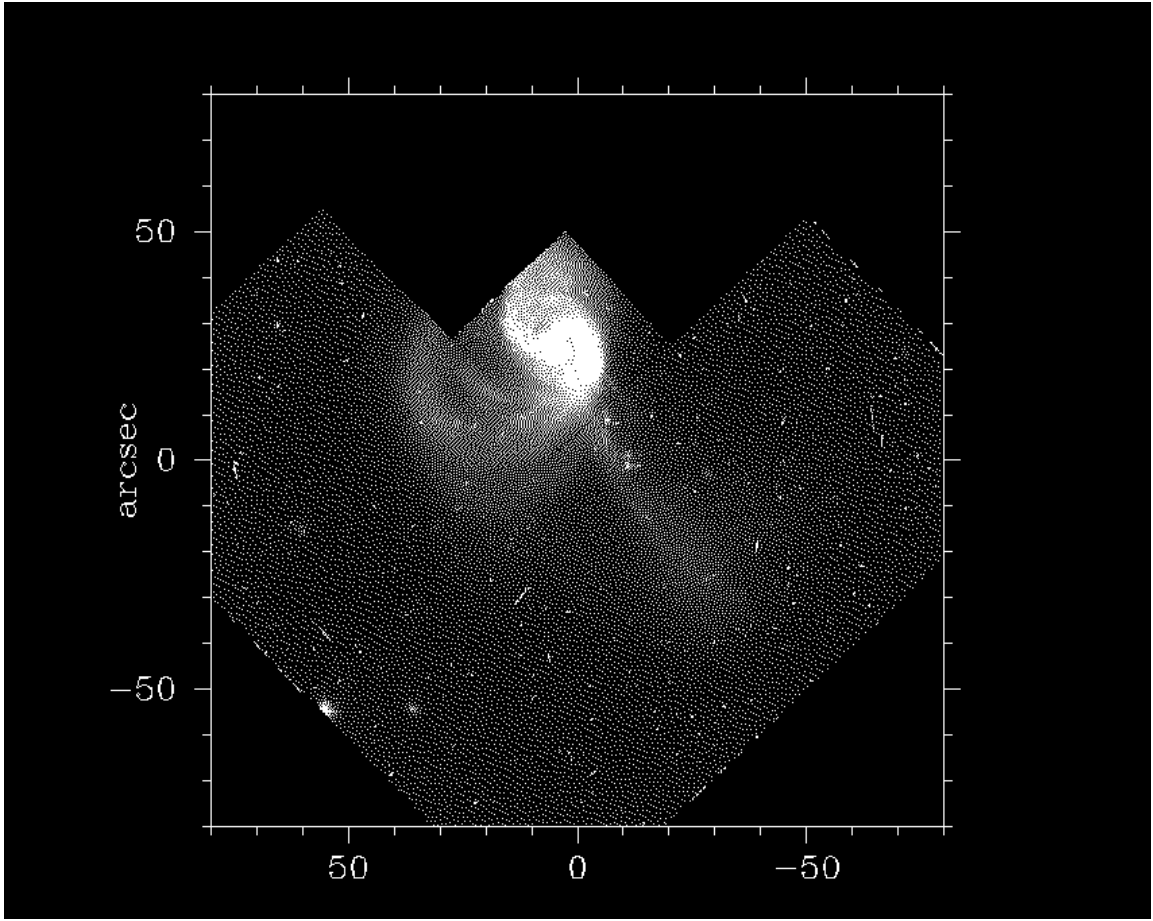}{425pt}{0}{80}{80}{-250}{-80}
\vspace{1cm}
\caption{An {\it HST}/WFPC2 F606W (optical) image of
NGC~1614 displayed on a logarithmic scale. 
The field of view is approximately 160\arcsec \ on a side. 
The orientation is north up, east to the left.}
\end{figure*}

\section{OBSERVATIONS}
\subsection{{\it HST}/NICMOS Observations}
{\it HST}/NICMOS observations of NGC~1614  were obtained in February
1998 using cameras NIC1 and NIC2. The pixel sizes are
0.043\arcsec\,pixel$^{-1}$ and 0.076\arcsec\,pixel$^{-1}$
respectively. Table~1 lists details of the observations.
Standard data reduction procedures were applied (see Alonso-Herrero
et al. 2000a for more details).
The flux calibration was performed using
the conversion factors based on measurements of the standard star P330-E
during the Servicing Mission Observatory Verification (SMOV) program.

The fully-reduced images were rotated to the usual 
orientation with north up, east to the left. They are shown in Figure~2. 
In addition, we constructed
an infrared color map using the NIC2 F160W and NIC2 F222M images,
which is equivalent to a ground-based $H-K$ color map.
In the NIC1 F110M image (highest angular resolution)
the nucleus of  NGC~1614 appears to be slightly resolved
with a FWHM of 0.15\arcsec \ (or 45\,pc for
the assumed distance), whereas for the unresolved
sources (the stellar clusters discussed in Section 3.2) we measured
FWHMs of approximately 0.11\arcsec \ (or 35 \,pc).

\begin{figure*}
\figurenum{2}
\plotfiddle{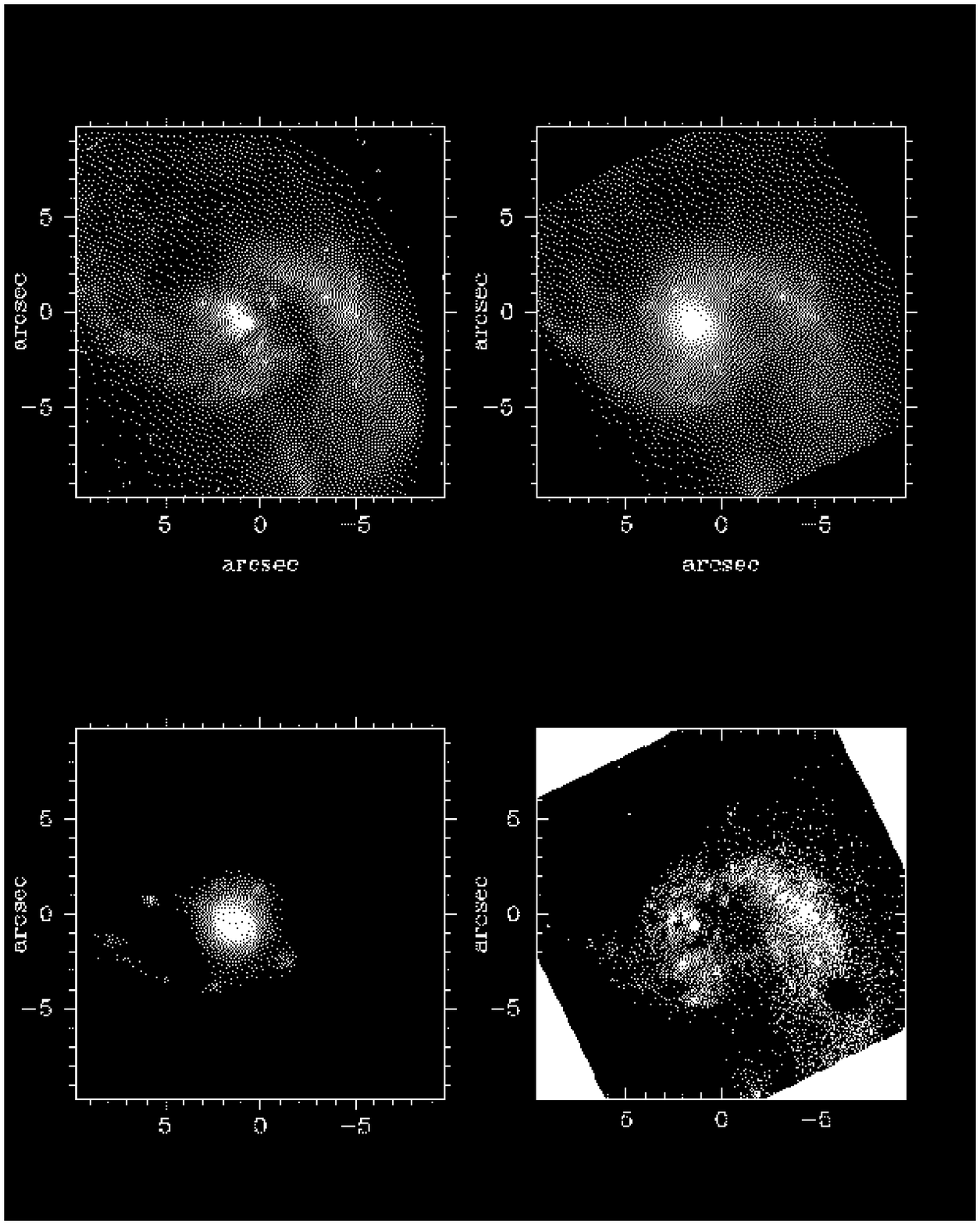}{425pt}{0}{70}{70}{-250}{-80}
\vspace{3cm}
\caption{From left upper corner clockwise: WFPC2 F606W 
(optical), NIC2 F160W (infrared $H$-band),
NIC2 F160W - F222M ($H-K$ color map),
and line emission (continuum subtracted) Pa$\alpha$ 
(at $\lambda_{\rm rest} = 1.87\,\mu$m) images.
The orientation is north up, east to the left. The field
of view is $19.5\arcsec \times 19.5\arcsec$. In the $H-K$ 
color map the dark colors indicate regions of higher extinction, 
with values of the $K$-band extinction ranging from $A_K = 
0.3\,$mag (white) to $A_K = 1.2\,$mag (black).}
\end{figure*}

The continuum subtracted Pa$\alpha$ image was produced by
a straight subtraction of the flux calibrated
NIC2 F187N image (continuum) from the
flux calibrated NIC2 F190N image (line+continuum).
NICMOS only provides narrow continuum bands to one side
of the emission line. For NGC~1614 the continuum image lies to the
blue, and therefore if some extinction is present the
continuum at the emission line wavelength may be slightly
over-subtracted. Table~2 gives the photometry of the
H\,{\sc ii} regions in the ring of star formation
and the brightest ones along the spiral arms.

\begin{figure*}
\figurenum{3}
\plotfiddle{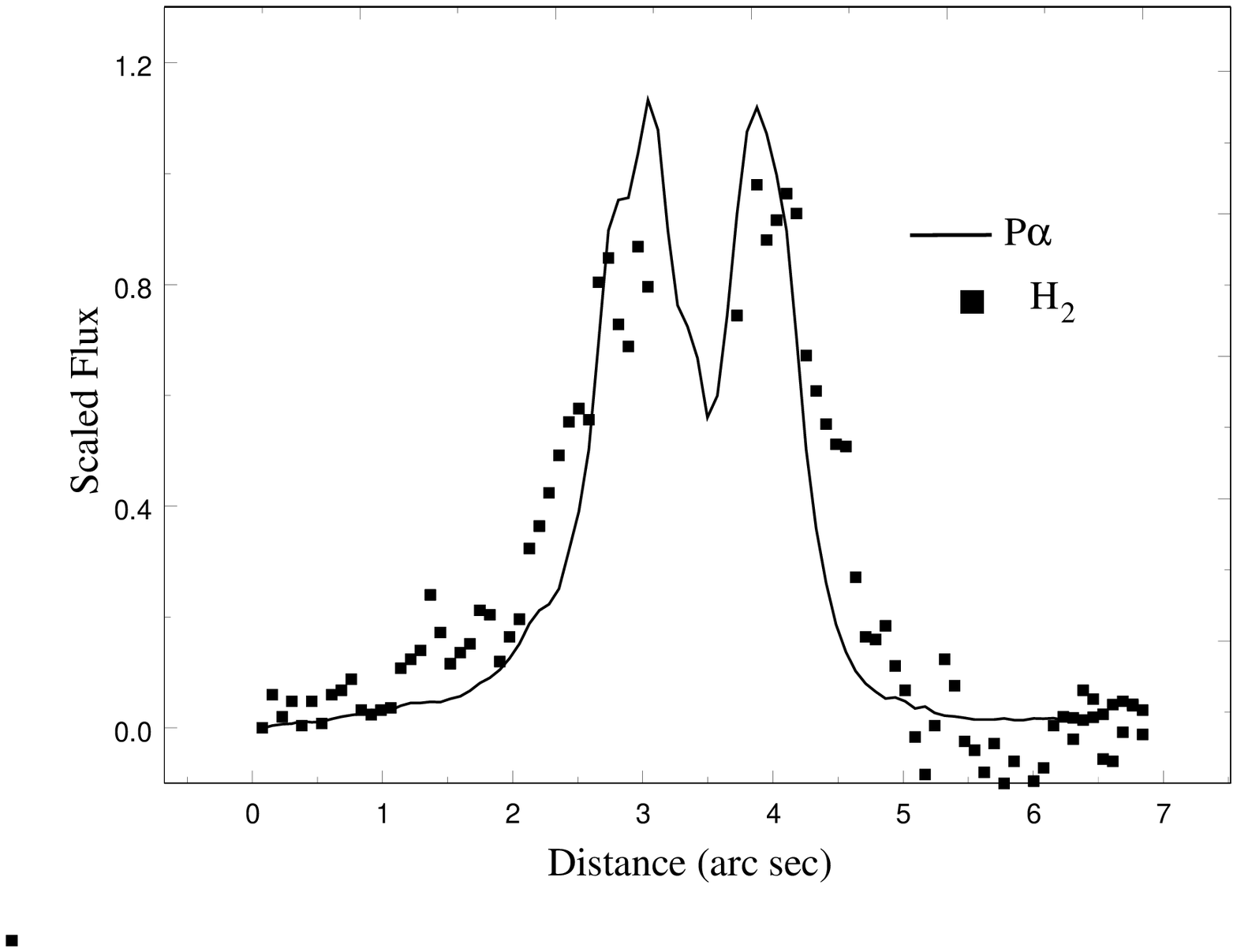}{425pt}{0}{70}{70}{-200}{180}\vspace{1cm}
\vspace{-8cm}
\caption{Comparison of the profile through the nucleus of 
the Pa$\alpha$ and (1,0)S(1) H$_2$ emission. The data 
were extracted from the narrow band NICMOS images in each line 
by averaging across the nucleus with a width of 0.61\arcsec\
oriented at position angle 170\arcdeg.}
\end{figure*}

A continuum subtracted H$_2$ image was produced in a similar
manner, using the F212N (continuum) and F215N (line+continuum) filters.
Because the H$_2$ line is relatively weak, the image
subtraction near the nucleus is not reliable. Artificial images
of point sources were generated for both filters and used
in experiments to determine where useful information could
be obtained. Based on these experiments, we have excluded
the data within 0.25\arcsec\ of the nucleus from further analysis. 
Figure~3 shows the surface brightness in Pa$\alpha$
compared with that in the (1,0) S(1) H$_2$ line at $\lambda_{\rm rest}
= 2.12\,\mu$m.
We have averaged the data along an artificial slit
0.61\arcsec\ in width and oriented at position angle
170\arcdeg\, selected to avoid the secondary nucleus
and the brightest H\,{\sc ii} regions in the
circumnuclear ring (see below). The H$_2$ extends
farther out from the nucleus than does
the Pa$\alpha$.

\begin{deluxetable}{lccccc}
\tablewidth{12cm}
\tablefontsize{\small}
\tablecaption{H\,{\sc ii} regions.}
\tablehead{\colhead{\#} &  \colhead{$\Delta$ R.A.} &
\colhead{$\Delta$Dec.} &
\colhead{$f({\rm Pa}\alpha)$} & \colhead{$\log L({\rm Pa}\alpha)$}
& \colhead{$\log N_{\rm Ly}$}\\
 & (s) & (\arcsec) & (erg cm$^{-2}$ s$^{-1}$) & (erg s$^{-1}$) &
(s$^{-1}$)}
\startdata
\multicolumn{6}{c}{ring H\,{\sc ii} regions}\\
\hline
  1 &  0.05 &  0.10 & $2.53\times 10^{-14}$ & 40.08 & 52.87\\
  2 &  0.03 & $-0.19$ & $2.43\times 10^{-14}$ & 40.07 & 52.86\\
  3 & $-0.04$ &  0.24 & $2.09\times 10^{-14}$ & 40.00 & 52.79\\
  4 &  0.00 &  0.40 &   $2.27\times 10^{-14}$ & 40.04 & 52.83\\
  5 &  0.01 & $-0.41$ & $2.21\times 10^{-14}$ & 40.03 & 52.82\\
total ring& \nodata & \nodata & $6.10 \times 10^{-13}$ & 41.47 & 54.30 \\
\hline
\multicolumn{6}{c}{spiral arm H\,{\sc ii} regions}\\
\hline
  6 & $-0.08$ &  1.93 & $2.45\times 10^{-15}$ & 39.07 & 51.86\\
  7 & $-0.19$ & $-1.91$ & $1.14\times 10^{-15}$ & 38.74 & 51.53\\
  8 &  0.05 & $-3.12$ & $9.35\times 10^{-16}$ & 38.65 & 51.44\\
  9 &  0.25 & $-2.82$ & $7.18\times 10^{-16}$ & 38.54 & 51.33\\
 10 &  0.30 &  1.34 & $2.75\times 10^{-15}$ & 39.12 & 51.91\\
 11 &  0.44 & $-0.81$ & $8.36\times 10^{-16}$ & 38.60 & 51.39\\
 12 &  0.57 &  2.13 & $6.99\times 10^{-16}$ & 38.52 & 51.31\\
 13 &  0.66 &  3.32 & $9.91\times 10^{-16}$ & 38.64 & 51.43\\
\enddata
\tablecomments{The Pa$\alpha$ photometry of the H\,{\sc ii}
regions is through a
0.30\arcsec-diameter aperture ($95\,$pc), whereas the total flux
of the ring is through a 2\arcsec-diameter aperture ($\simeq 620\,$pc).
These values have not been corrected for extinction.}
\end{deluxetable}

Alonso-Herrero et al. (2000a) discuss a
NICMOS CO photometric index (defined as
${\rm CO}_{\rm NICMOS} = \frac{f({\rm F222M}) - f({\rm F237M})}
{f({\rm F222M})}$, where the fluxes $f$(F222M) and 
$f$(F237M) are in Jy). The spatial profile of
this index is shown in  Figure~4 together 
with the Pa$\alpha$ line emission (upper panel) and the 
NIC2 F222M ($K$-band) brightness profile
(lower panel). The errors in the CO index
account for the background subtraction uncertainties in both the 
NIC2 F222M and NIC2 F237M filters. The nominal errors of the Pa$\alpha$ 
and NIC2 F222M profiles are smaller than the size of the symbols.

\begin{figure*}
\figurenum{4}
\plotfiddle{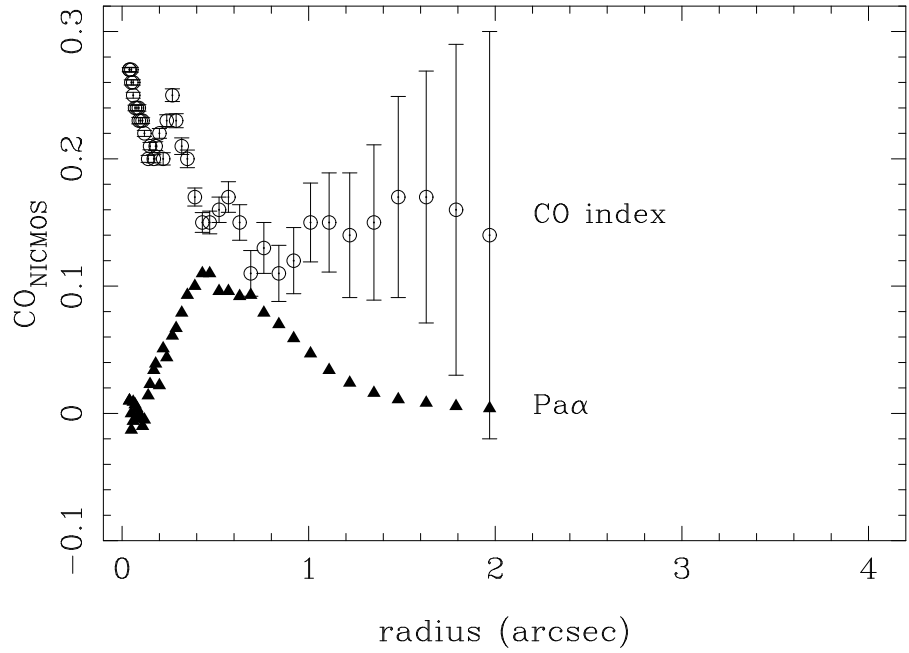}{425pt}{270}{70}{70}{-200}{550}
\plotfiddle{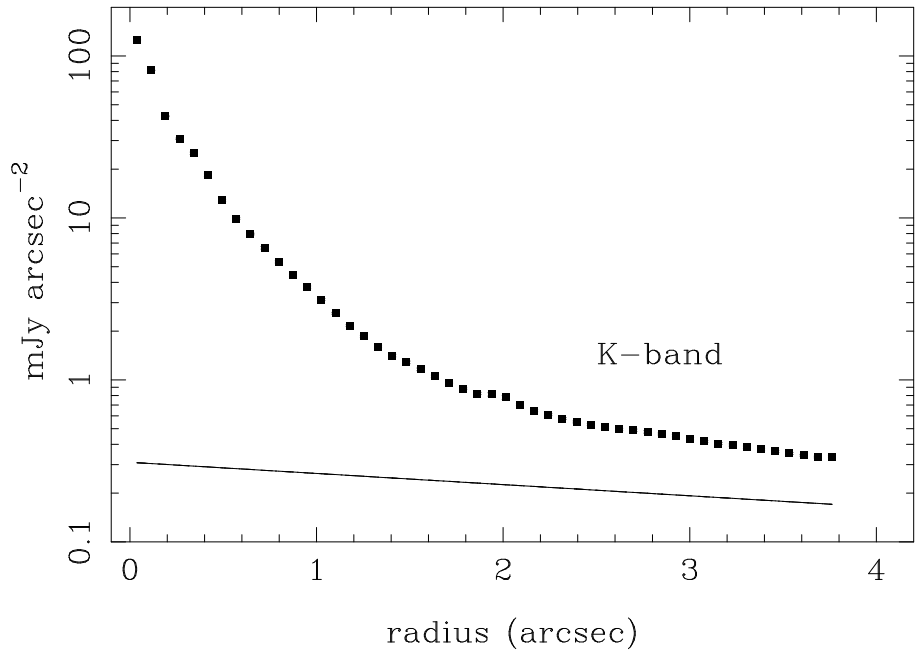}{425pt}{270}{70}{70}{-200}{650}
\vspace{-7cm}
\caption{{\it Upper panel}: CO$_{\rm NICMOS}$ surface 
brightness profile. 
For comparison we also plot the Pa$\alpha$ profile (in arbitrary units) 
to show the location  of the ring of star formation. {\it Lower panel}: 
NIC2 F222M ($K$-band) surface brightness profile. The solid line is 
the fit to an exponential disk (see Section~3.3.2 and Figure~7).}
\end{figure*}

\subsection{{\it HST} archival data}

A Wide Field Planetary Camera 2 (WFPC2)
image of NGC~1614 was retrieved from the {\it HST}
data archive. This image was taken through the
F606W filter in November 1994.  Standard pipeline reduction
procedures were applied. Because only a single exposure
was taken for this galaxy, the cosmic
rays are particularly difficult to remove,
especially the extended ones. The image is
saturated in the center. We rotated the cleaned
image to the usual orientation (see Figure~1).

\subsection{Near-Infrared Spectroscopy}

Near-infrared spectra were obtained at Steward Observatory's 2.3-m Bok
Telescope on Kitt Peak using FSpec (Williams et al.\ 1993) with 
an orientation ${\rm P.A}=90\arcdeg$.  The data
were reduced as described by Engelbracht et al.\ (1998).  The spectra
were extracted from a $2.4\arcsec\times6.0\arcsec$ region centered on
the galaxy nucleus and were flux-calibrated using ground-based imaging
data. The $J$-band spectrum at a resolution of
$\sim800$ is shown in Figure~5 (upper panel), while
spectra in the $H$ and $K$ bands at a resolution of $\sim3000$ are
shown in Figure~5 (middle and lower panels respectively). We have used 
the procedures discussed in Engelbracht et al. (1998) to
subtract a stellar continuum, so that accurate measurements
can be made of faint lines that might otherwise be affected
by stellar absorption features. In Table~3
we present emission line fluxes. Our detections of
the (1-0)S(0) and (2-1)S(1) lines of H$_2$ are supported
by weak detections of the same lines by Puxley \& Brand (1994).

The spectra also allow us to determine the CO index to
quantify the depth of the first overtone absorption bands
at 2.3$\mu$m. Using the method and
calibration discussed by Engelbracht et al.\ (1998), the 
CO index is 0.23, substantially stronger than for 
normal old stellar populations found in the nuclei
of non-starbursting galaxies. Our determination that the
CO absorption is stronger than expected for a "normal"
galaxy nucleus is in agreement with previous work (Ridgway,
Wynn-Williams, \& Becklin 1994; Puxley \& Brand 1994 as discussed
by Goldader et al. 1997).

\begin{deluxetable}{lcc}
\tablewidth{8cm}
\tablefontsize{\small}
\tablecaption{Line Flux Measurements\label{lineflux}}
\tablehead{\colhead{$\lambda_{vac}$} & \colhead{species} &
\colhead{flux}\\
\colhead{(\micron)} & \colhead{} & \colhead{($10^{-14}$ erg s$^{-1}$
cm$^{-2}$)}}
\startdata
1.0941 & Pa$\gamma$     &  8.2 $\pm$ 0.6 \\
1.2570 & [\ion{Fe}{2}]  &  3.9 $\pm$ 0.2 \\
1.2822 & Pa$\beta$      & 13.4 $\pm$ 0.3 \\
1.5339 & [\ion{Fe}{2}]  &  0.5 $\pm$ 0.2 \\
1.6114 & Br13           &  0.9 $\pm$ 0.3 \\
1.6412 & Br12           &  0.7 $\pm$ 0.2 \\
1.6440 & [\ion{Fe}{2}]  &  4.4 $\pm$ 0.2 \\
1.6774 & [\ion{Fe}{2}]  &  0.5 $\pm$ 0.2 \\
1.6811 & Br11           &  1.7 $\pm$ 0.3 \\
1.7367 & Br10           &  1.6 $\pm$ 0.1 \\
2.0338 & H$_2$(1,0)S(2) &  0.3 $\pm$ 0.1 \\
2.0587 & HeI            &  2.0 $\pm$ 0.1 \\
2.0735 & H$_2$(2,1)S(3) &  0.3 $\pm$ 0.1 \\
2.1218 & H$_2$(1,0)S(1) &  1.0 $\pm$ 0.1 \\
2.1542 & H$_2$(2,1)S(2) &  $<0.2$ \\
2.1661 & Br$\gamma$     &  4.0 $\pm$ 0.1 \\
2.2233 & H$_2$(1,0)S(0) &  0.3 $\pm$ 0.1 \\
2.2477 & H$_2$(2,1)S(1) &  0.2 $\pm$ 0.03 \\
\enddata 
\end{deluxetable}

\begin{figure*}
\figurenum{5}
\plotfiddle{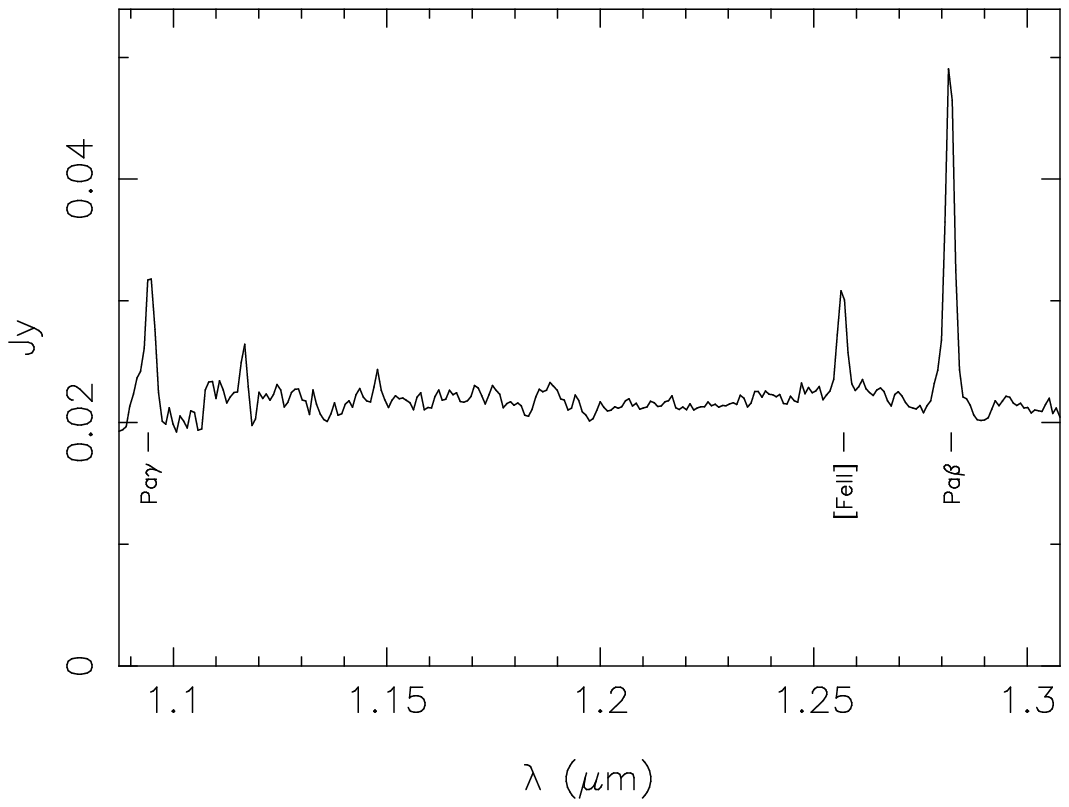}{425pt}{0}{80}{80}{-180}{200}
\plotfiddle{figure5b.ps}{425pt}{270}{40}{40}{-180}{670}
\plotfiddle{figure5c.ps}{425pt}{270}{40}{40}{-180}{900}
\vspace{-23cm}
\caption{Flux calibrated 
$J$-band spectrum (upper panel), $H$-band spectrum (middle panel)
and $K$-band spectrum (lower panel). The lower spectra in the
panels for $H$ and $K$ are after subtraction of a stellar continuum
and show the relative emission line strengths more clearly.}
\end{figure*}

\section{DISCUSSION}

\subsection{The extinction}
Before we interpret the properties of NGC~1614, we need
to apply an extinction correction. The extinction to the nucleus of
NGC~1614 has been estimated a number of
times. Some works (e.g., Neff et al. 1990; Shier, Rieke, 
Rieke 1996) have found
relatively small values for the visual
extinction ($A_V = 3-5\,$mag with a foreground screen model)
from optical and infrared colors, and infrared emission lines.
Puxley \& Brand (1994) used
optical and near infrared hydrogen recombination lines, and claimed that
the extinction to NGC~1614 could not
be modeled with a simple foreground screen model. Instead they used
a composite model (mixture of dust and gas, and a foreground
screen) and inferred a total extinction in the optical of
$A_V \simeq 11\,$mag. Ho, Beck, \& Turner (1990) derived a
relatively large extinction between Br$\gamma$ and Br$\alpha$,
but their result is subject to significant calibration uncertainties
(see also Puxley \& Brand 1994).

\begin{deluxetable}{lccccc}
\tablewidth{10cm}
\tablefontsize{\small}
\tablecaption{Cluster Photometry in the $H$-band.}
\tablehead{\colhead{\#} &  \colhead{$\Delta$ R.A.} &
\colhead{$\Delta$Dec.} &
\colhead{$m_{\rm F160W}$} & \colhead{$m_{\rm 1.60} - m_{\rm 2.22}$}
& \colhead{$M_H$}\\
& (s) & (\arcsec)}
\startdata
1  & $-0.47$ & $-0.66$ &   18.7 &0.56 & $-15.4$\\
2  & $-0.43$ & $-1.82$ &   19.1 &0.32 & $-15.0$\\
3  & $-0.42$ & $-3.49$ &   18.9 &0.52 & $-15.2$\\
4  & $-0.42$ &  0.34   &   17.5 &0.44 & $-16.6$\\
5  & $-0.31$ &  1.45   &   16.9 &0.38 & $-17.2$\\
6  & $-0.27$ &  2.42   &   18.8 &0.05 & $-15.3$\\
7  & $-0.21$ & $-8.71$ &   18.6 &0.36 & $-15.4$\\
8  & $-0.16$ &  2.41   &   18.2 &0.69 & $-15.8$\\
9  & $-0.13$ &  2.04   &   20.0 &1.94 & $-14.0$\\
10 & $-0.10$ &  1.37   &   17.9 &0.28 & $-16.1$\\
11 & $-0.01$ &  2.81   &   18.9 &0.39 & $-15.1$\\
12 & $-0.08$ & $-2.23$ &   17.5 &0.47 & $-16.5$\\
13 &  0.07   &  1.70   &   17.0 &0.84 & $-17.0$\\
14 &  0.16   &  1.04   &   18.1 &0.70 & $-15.9$\\
15 &  0.30   &  1.44   &   19.3 &1.18 & $-14.8$\\
16 &  0.52   & $-0.08$ &   18.9 &0.52 & $-15.1$\\

\enddata
\tablecomments{Cluster photometry is through a 0.61\arcsec-diameter
aperture (corrected for aperture effect). The $m_{1.60}-m_{2.22}$ color is 
similar to a ground-based $H-K$ color.}
\end{deluxetable}

Using the fluxes of the two [Fe\,{\sc ii}] emission lines 
at $1.257\,\mu$m and 
$1.644\,\mu$m (whose ratio is largely independent of nebular
conditions) and the hydrogen recombination
lines Pa$\beta$ and Br$\gamma$ and a foreground dust 
screen model,  we find values of the extinction
to the gas in the $K$-band of $A_K=0.40-0.49\,$mag,
corresponding to $A_V \sim 4\,$mag.
A comparison of dust models (from Witt, Thronson,
\& Capuano 1992) with the simple foreground screen model 
(c.f., the analysis of NGC~253 by Engelbracht et al. 1998) shows 
that the correction needed for all the 
infrared lines ([Fe\,{\sc ii}]$1.26\,\mu$m,
[Fe\,{\sc ii}]$1.64\,\mu$m, Pa$\beta$ and Br$\gamma$)
and the broad-band colors is similar. Thus a  foreground
dust screen model (with extinction in the $K$-band of
$A_K= 0.4-0.5\,$mag) is a good approximation for
the near-infrared extinction to the ionized gas.
This approximation may not give the correct
extinction  in the optical, but we do not
use any optical data for our study of the star formation properties of
this galaxy.
                 
The infrared $H-K$ color map (Figure~2 and also the close-up
in Figure~6) shows that the actual distribution of the
extinction is quite patchy. If we assume an intrinsic
color $H-K=0.25$ for the stellar population and the
Rieke \& Lebofsky (1985) extinction law, we find values
of the $K$-band extinction to the stars of $A_K=0.3-0.4\,$mag
to the east of the center of NGC~1614, whereas the $K$-band
extinction to the west is higher, ranging between approximately
0.2 and 1.2\,mag. The $H-K$ color of the nucleus (through
a 0.76\arcsec-diameter aperture) is 0.41, which would imply a $K$-band
extinction of $A_K=0.3\,$mag.

\begin{figure*}
\figurenum{6}
\plotfiddle{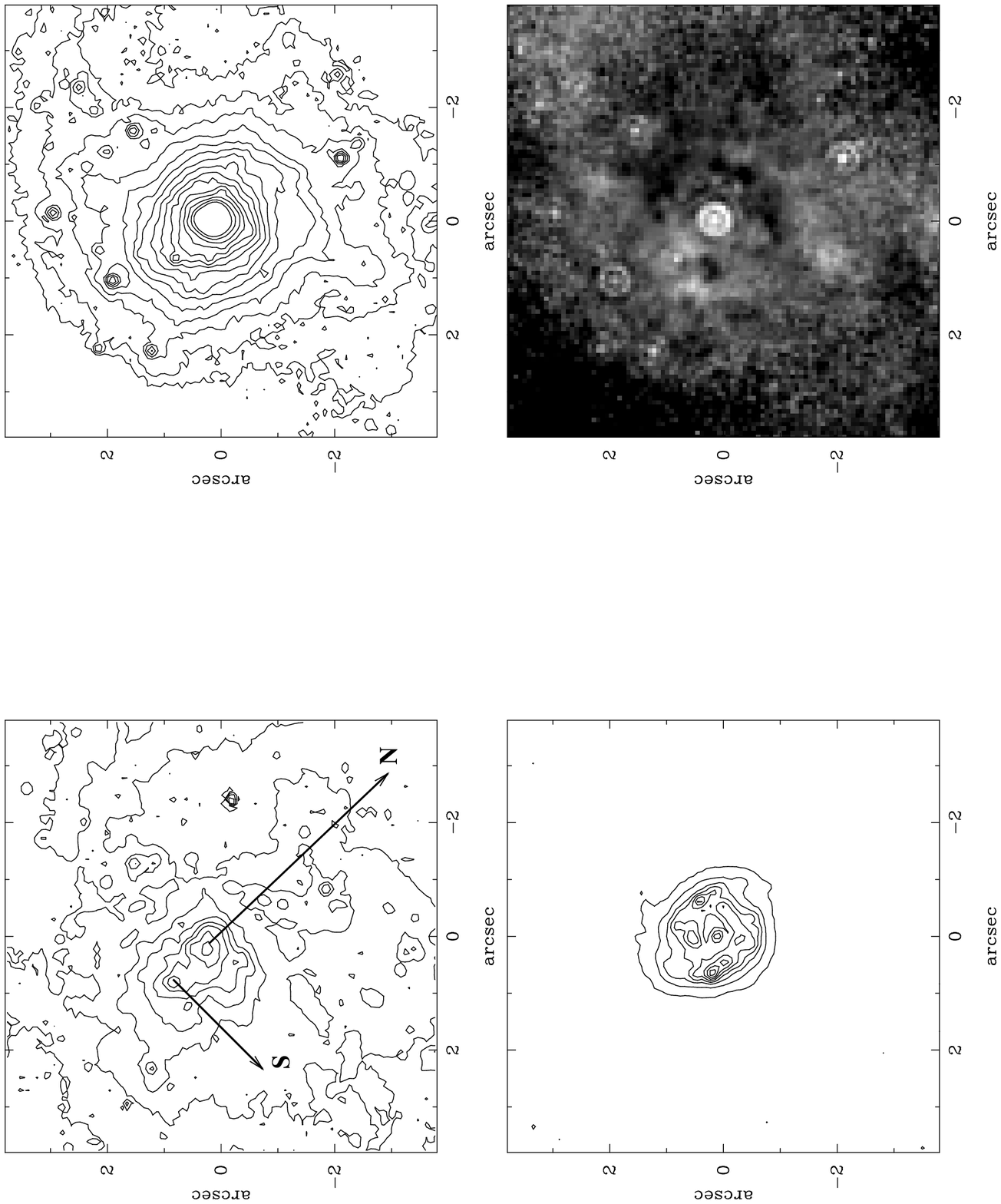}{425pt}{270}{70}{70}{-280}{500}
\vspace{-2cm}
\caption{Close ups of the central $8\arcsec \times 8\arcsec$.
From the top left panel and clockwise:
contour plot on a linear scale in the WFPC2 F606W filter,
contour plot on a linear scale
in the $H$-band (NIC2 F160W image), grey-scale $H-K$ color map
(dark means regions with higher extinction, grey
scale as in Figure~2) and
contour plot of the continuum-subtracted
emission line Pa$\alpha$ (NIC2 F190N - NIC2 F187N). The orientation
for all the images is north up, east to the left. We indicate 
the position of the nucleus (N) and the secondary nucleus (S) in 
the WFPC2 F606W panel.}
\end{figure*}

Compared with other dusty starbursts, the levels of extinction
in NGC~1614 are low. We can determine the near infrared
properties of the nucleus of NGC~1614
reasonably accurately by taking an average extinction of
$A_K = 0.5\,$mag and making a first order correction by
assuming a simple foreground screen of dust.
This value for the extinction agrees well with
foreground screen model of Puxley \& Brand (1994), $A_K = 0.43\,$mag.
They show that an underestimate of the Ly continuum by
about 20\% may result compared with more sophisticated
models. However, from the combination of our slightly higher
assumed extinction level and other minor differences,
our derived Ly continuum is similar to the value they feel is most accurate. 
In general, if there are regions of very heavy extinction, or where the
extinction is optically thick at the wavelengths of our
observations, then we would have
underestimated the starburst parameters. Our conclusions
about the galaxy would become stronger with this error corrected.

The relatively modest extinction in a face-on starburst is an important
advantage for studies of NGC~1614. The foundation for much of
this paper is the resulting possibility to determine the
starburst properties more reliably than for most
other well-studied examples.

\subsection{Morphology}

The NICMOS images provide important new insights to the
morphology of the starburst in NGC 1614. Figure~2
presents from the upper left corner and clockwise, the optical
WFPC2 F606W image, NIC2 F160W ($H$-band), NIC2 F160W - NIC2 F222M
($H-K$ color map), and continuum subtracted Pa$\alpha$ image. The
field of view of these images is $19.2\arcsec \times 19.2\arcsec$.
Figure~6 presents close-ups of the central regions ($8\arcsec\times
8\arcsec \simeq 2.5\,{\rm kpc}\times 2.5\,{\rm kpc}$)
as in Figure~2.

The nucleus of NGC~1614 is extremely luminous, with an
$H$-band absolute magnitude measured through a 0.76\arcsec-diameter aperture
of $M_H=-23\,$mag (corrected for
extinction). The bright infrared nucleus is surrounded
by a number of fainter sources and similar sources lie along
the spiral arms. These objects are probably
stellar clusters. We measured $H$-band luminosities (see Table~4)
ranging up  to $M_H = -17.2\,$mag for the brightest sixteen clusters
(excluding those located in the vicinity of the bright nucleus),
similar to the cluster luminosities measured in Arp~299
(Alonso-Herrero et al. 2000a) and other luminous and
ultraluminous infrared galaxies (Scoville et al. 2000).
Neff et al. (1990) noted the presence of a knot some 10\arcsec \
SW of the nucleus in a $B$-band image
(figure~2 in their paper), and suggested that it could be interpreted
to be fragments of a second galaxy. This knot (tentatively
identified with cluster \#7,
Table~4) is on the edge of
our NIC2 F160W image. It is most likely to be 
another stellar cluster from its observed luminosity.

Ground-based imaging shows the nucleus of the galaxy
to be elongated at position angle 39$\arcdeg$ (e.g., Mazzarella
\& Boroson 1993). The WFPC2 F606W image resolves the
central 2\arcsec\ of NGC~1614 into two prominent sources at
a similar position angle (see Figure~6).
The brighter of these sources coincides with the infrared nucleus.
The other optical source lies at the position of
a secondary peak in the infrared images to the NE of the nucleus, but
it is far less prominent in the infrared than
in the F606W image. Its $H$-band luminosity
is about 30 times fainter than the primary nucleus
($M_H = -18.5\pm0.4\,$mag; the error accounts for the
uncertainty in the determination of the underlying background
because of its proximity to the bright nucleus).
The $H$-band absolute magnitude of the secondary peak is a
magnitude or more brighter
than the brightest of the stellar clusters, suggesting
that it is the nucleus of an interacting small galaxy.
The small extinction toward the secondary nucleus,
as indicated by its blue color from optical to infrared, places
it in front of the main nucleus and the activity around it (see 
Figure~6).

The CO band in the primary nucleus is very
strong as shown by both our spectroscopy (Section~2.3) and imaging
(Figure~4), much stronger than expected for an old stellar
population. The most plausible explanation for this
behavior is that the light from this nucleus is dominated
by red supergiants that are the product of a starburst
of age 10 or more million years. As shown in Figure~4,
outside the starburst ring  the CO index assumes
a value appropriate to an old stellar
population, presumably that of the underlying galaxy.

The Pa$\alpha$ line emission image reveals a
nuclear ring of star formation with an
approximate diameter of 650\,pc. The Pa$\alpha$ morphology is remarkably
similar to the 6\,cm radio map (Neff et al. 1990). Although
the two available mid-infrared images (Keto et al. 1992;
Miles et al. 1996) do not agree in detail, they do
show that the emission in the central
part of the galaxy is confined to a 2" diameter
region. The image of Miles et al. (1996),
which should have higher angular resolution and signal
to noise, suggests that the emission arises from
the ring seen in Pa$\alpha$.

Comparison of small aperture ($\sim$ 5\arcsec\
diameter) ground-based photometry (Rieke \& Low 1972;
Lebofsky \& Rieke 1979; Carico et al. 1988) with the
IRAS fluxes (in a 45\arcsec\ by 250\arcsec\ beam)
shows that the infrared emission of the entire galaxy
is strongly concentrated into the central region.
In fact, if approximate corrections
are made from the ground-based to IRAS effective wavelengths
and for the wavelength weighting introduced in IRAS to
reduce color corrections to the fluxes, more than 80\% of
the total infrared flux at 12 $\mu$m originates in the nucleus, and
hence, from the Miles et al. (1996) image, primarily in the
circumnuclear ring. At 25$\mu$m, making similar corrections
to the measurement by Lebofsky \& Rieke (1979), we find
about 75\% of the total flux to originate in the nucleus.
Given the relation between far infrared and radio fluxes,
we can also estimate the central concentration in the
far infrared from radio data. We compare the survey results
by Griffith et al. (1995) in a 4.2 arcmin beam at 6cm with
the VLA map of the nucleus reported by Neff et al.
(1990). We have corrected the flux scale in the latter
reference upward by 11\%, as they suggest. We
find that $\sim$ 60\% of the 6cm flux lies in the
nuclear ring. These three estimates are consistent in 
stating that about 70\% of the far infrared luminosity 
arises from the nuclear region. In the following, we can 
ascribe most of the far infrared luminosity to the circumnuclear
ring of H\,{\sc ii} regions. 

The integrated far infrared luminosity indicates a star
formation rate of 52 M$_\odot$ yr$^{-1}$ from equation(3) of
Kennicutt (1998). If 70\% of this star formation occurs
in the nucleus, we predict a level there of 36 M$_\odot$ yr$^{-1}$. 
The Pa$\alpha$ luminosity
in the circumnuclear ring (see Table~2) indicates,
after correction for reddening, a star formation rate of
$\sim$ 36 M$_\odot$ yr$^{-1}$ from equation (2) of Kennicutt (1998),
in perfect agreement. 

The H\,{\sc ii} regions in the ring of NGC~1614 are extremely luminous with
an equivalent ionizing luminosity an order of magnitude
greater than that of 30 Doradus.
A similar rich population of bright H\,{\sc ii} regions has been
found in the luminous starburst in Arp~299
(see Alonso-Herrero et al. 2000a). The H\,{\sc ii} regions in
the spiral arms are weaker than those in the ring,
but still exceedingly luminous. Those listed in Table~2
are all similar in luminosity to 30 Dor (for which log L$_{Pa\alpha}$ $\sim
38.8\,$erg s$^{-1}$; Kennicutt, Edgar, \& Hodge  1989).
30 Dor is usually considered the prototype {\it super}-H\,{\sc ii} region.
Such objects are very uncommon in normal galaxies (e.g., Kennicutt et
al. 1989; Rozas, Beckman, \& Knapen 1996). The existence
of so many comparable to 30 Dor and even with greater
ionizing luminosity appears to be associated with strong
starbursts (see Alonso-Herrero et al. 2000a).

Just outside the ring of H\,{\sc ii} regions, the $H-K$ extinction
map shows a partial ring of high extinction. This ring is
obliterated to the NE due to the overlying secondary nucleus,
but otherwise it can be seen for more than 220$\arcdeg$. It
traces the dense interstellar molecular material as discussed
in the following section.

The molecular hydrogen appears to extend from the ring of
H\,{\sc ii} regions into the molecular cloud 
(see Figure~3). The detection of the (2-1) S(3) line at 
$2.0735\,\mu$m and the (2-1) S(1) line at $2.2471\,\mu$m 
(the latter line both in our spectrum and in that
of Puxley \& Brand 1994) suggests that a significant
fraction of the emission is due to fluorescence in
low density gas, for which the ratio of these lines to
the $2.12\,\mu$m line is $\sim 0.5$ (e.g. Black \& van Dishoek 1987;
Sternberg \& Dalgarno 1989). The relative H$_2$ line
strengths are very similar to those in NGC 253 (Engelbracht et al.
1998), for which we derived that the excitation was
probably shared between fluorescence and shocks, with nearly
2/3 of the total near infrared H$_2$ luminosity from the
former mechanism. The morphology suggests that at least some of the H$_2$ originates at a shock penetrating into the molecular ring from
the H\,{\sc ii} ring. Spectra of higher signal to
noise and angular resolution will be needed to sort
out the excitation conditions for the H$_2$ in more detail.

Figure~7 summarizes the distribution of starlight on a
large scale. We have used ellipse fitting to generate
surface brightness profiles from the NICMOS $K$-band image
(Figure~4) and from a ground-based image (Shier et al. 1996).
These profiles join smoothly and trace the surface brightness
to 20\arcsec\ radius ($> 6 \,$kpc). The galaxy is dominated
by an exponential disk with a scale length of
6.3\arcsec\ ($1.95\,$kpc) from a radius of 5\arcsec\ ($1.6\,$kpc)
outward. Within a radius of 3\arcsec ($1\,$kpc), there is a
bright central source fitted well by an r$^{1/4}$ law
with a scale length of 1.14\arcsec \ or 350\,pc.
At least in the innermost arcsec this region is dominated
by the output of the newly formed stars and not by a
traditional bulge of old stars.

\begin{figure*}
\figurenum{7}
\plotfiddle{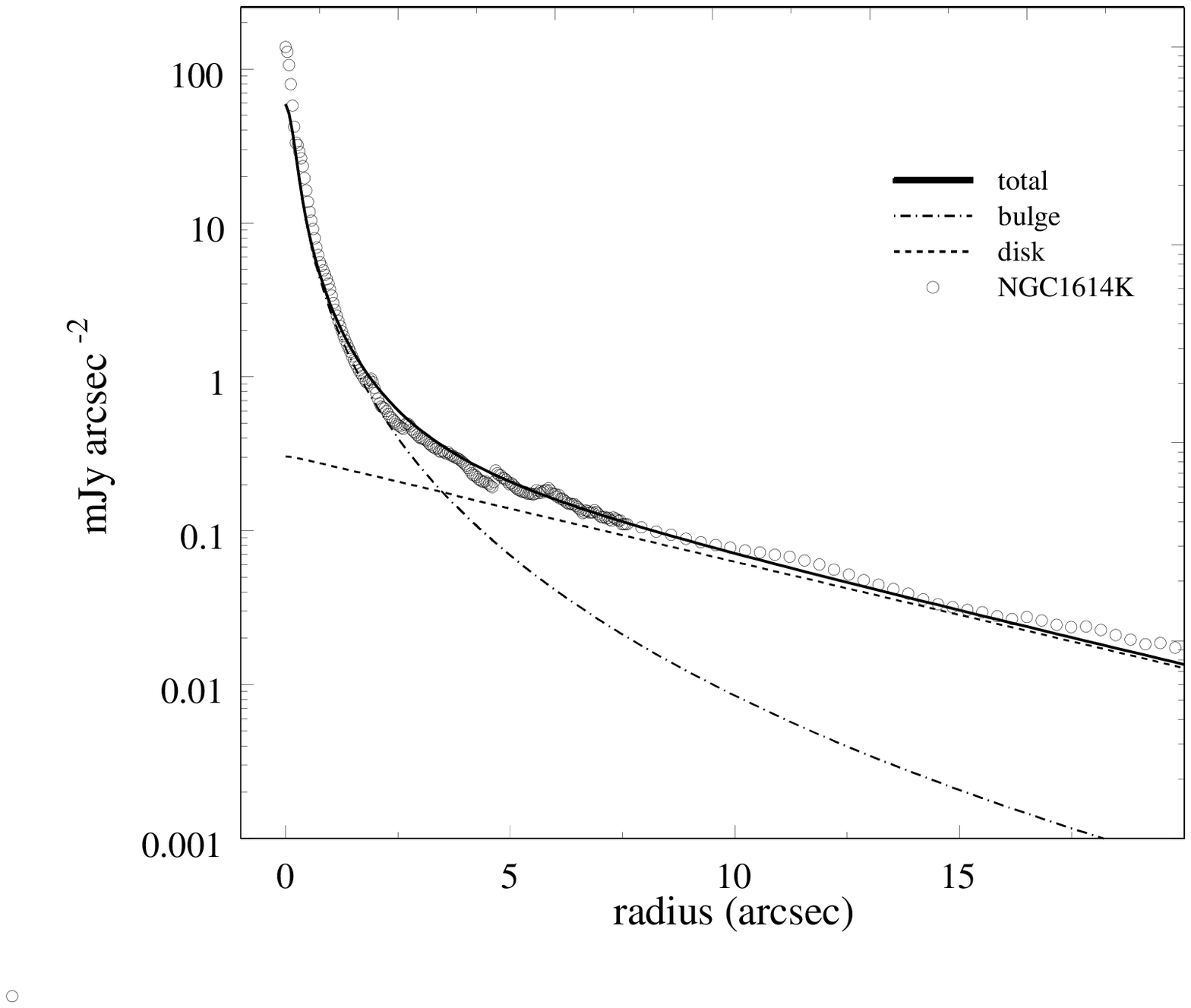}{425pt}{0}{70}{70}{-200}{100}
\vspace{-4cm}
\caption{Surface brightness profile at $K$, combining 
NICMOS and ground-based images. The profile has been
fitted with a combination of an exponential for the disk
and an r$^{1/4}$ law for the central region.}
\end{figure*}

Combining the results reported above,
NGC~1614 appears to be a textbook example of a propagating
starburst that started in the nucleus of a late type, large
spiral galaxy, has grown outward to a radius of
$\sim$ 300 pc, and is potentially
still growing into a circumnuclear nuclear ring of molecular
material just outside this radius (see also discussion
in Section~3.4). This picture will be made
more quantitative in a following section where we carry out
evolutionary synthesis modeling of the starburst.
Our data re-emphasize the result of Neff et al. (1990) that
the activity in this galaxy is dominated by star formation
and not by an AGN -- there is no compact nuclear source either
in the near or mid infrared, the emission lines are
all narrow, and correlations observed in star
forming regions such as the ratio of hydrogen recombination lines
to far infrared luminosity appear to hold in NGC 1614. An interesting
additional result is that the starburst is accompanied by
a number of {\it super}-H\,{\sc ii} regions in the spiral
arms of the galaxy. The presence of a secondary nucleus
out of the plane of NGC~1614 suggests that all this activity
has been triggered by an interaction with a smaller galaxy,
which has by now largely been destroyed. There
is further evidence for a merger from the tidal tails in the outer
structure of NGC~1614 (Figure~1).

\subsection{Mass}

An aperture of diameter 2\arcsec \ (620\,pc at the galaxy distance)
includes the starburst and molecular
ring. To understand the behavior in these regions, we need
to determine the total mass and the mass of various
constituents within this aperture. We first estimate the
dynamical mass and then show that it is virtually all
accounted for by the population of old, pre-starburst stars.
The mass budget leaves very little for the starburst and
molecular gas, as summarized in Table~5. This budget is
strongly incompatible with the standard
conversion of CO to gas mass and also challenges starburst
models using conventional forms of the initial mass function.

\begin{deluxetable}{lcc}
\tablewidth{9cm}
\tablefontsize{\small}
\tablecaption{Mass Budget Summary for the nuclear ring.}
\tablehead{\colhead{Component} &  \colhead{Estimate*} & \colhead{Adopted*}}
\startdata
Total dynamical mass  &    & 13 \\
$-$ From stellar velocity dispersion  & 12 &  \\
$-$ From Pa$\alpha$ ring rotation  & 14 &  \\
Mass in old stars  &     & 12 \\
Modeled starburst mass  &   & 5.5 \\
Molecular gas mass  &   & 3 \\
$-$ From standard CO conversion  & 48 &  \\
$-$ From extinction  & $\ge 1-2$ &  \\
$-$ From Toomre Q condition  & $\le 4$ & \\
$-$ For optically thin CO  & 2  &  \\

\enddata
\tablecomments{* units of $10^8\,{\rm M}_\odot$}
\end{deluxetable}

\subsubsection{Dynamical mass}

Shier et al. (1996) modeled the velocity dispersion
measured in the 2.3$\mu$m CO band head to derive the dynamical
mass for a 1000\,pc (3\arcsec) diameter region in NGC 1614.
They considered three models for the stellar
distribution: two bulge models and
a disk model. The mass was only weakly dependent on
which model was used for its derivation, although the
mass in the disk model would be about twice that in the
bulge ones if the disk were assumed to be at the inclination
we find for the circumnuclear ring. However, since the $K$-band
surface brightness profile of the inner 2-arcsec radius is well fitted
with an $r^{1/4}$ law (see Figure~7), here we adopt the
highest value for the spherical geometry (bulge) models
of Shier et al. (1996), 1.8 $\times$ 10$^9$ M$_\odot$, and
make an approximate correction in proportion to diameter
to the 2\arcsec \ diameter region which is thus indicated
to have a mass of 12 $\pm$ 4 $\times$ 10$^8$ M$_\odot$.

An independent estimate of the mass in the
central region of NGC~1614 can be obtained using the Br$\gamma$ spectroscopy
presented in Puxley \& Brand (1999), and our knowledge of the
morphology and size of the emitting region (a ring).
It is expected that the ring is associated with a Lindblad resonance
or other similar dynamical feature and is close to round.
Puxley \& Brand (1999) detected
two peaks of line emission separated by $151\,$km s$^{-1}$.
We have obtained the 
ring rotational velocity by modeling the line
profile expected for a rotating ring, convolved
with a Gaussian to represent both the instrumental
profile (resolution 39 km s$^{-1}$) and the dispersion within the beam. We find
models with dispersions between 38 and 64
km s$^{-1}$ fit equally well as the double Gaussian
used by Puxley \& Brand. In all the satisfactory models,
the peak rotational velocity is very close to 1.4 times the 
velocity that would have been deduced just from
the separation of the Gaussians fitted by Puxley and Brand. That is,
the rotational velocity is 106\,km s$^{-1}$, uncorrected
for inclination. The total rotational velocity is determined
by de-projecting for an inclination
of $i=51\arcdeg$ (determined from the observed ellipticity
and an assumption that the ring is round). 

From the $H$-band morphology we know that the mass is
symmetrically distributed inside the H\,{\sc ii} ring,
and therefore we can use the virial theorem in its simplest form
to convert the rotational velocity to a mass. The conversion depends 
weakly on the mass distribution. Given the good fit of the r$^{1/4}$ 
model to the starlight in the center of the galaxy (see Section 3.2), 
coupled with the large mass represented by these stars (see next section), 
we have assumed a spherically symmetric mass distribution. We find that 
the total mass in a region with diameter 2\arcsec\
$\sim\,$ 620 pc is $14\times 10^8\,{\rm M}_\odot$. The mass for a 
spherically symmetric distribution lies midway between the minimum 
possible value, for a thin disk, and the maximum, for a thin and narrow 
ring. The total range between these two extremes is a factor of two 
(i.e., $\sim$ $\pm 40\%$). Both because the situations at the 
extremes tend to be unstable and because of the observational evidence 
for a massive stellar spheroid within the ring, the systematic errors in
 our mass estimate should be much smaller than this range. 

For the following we adopt the average of the
two measures of $\sim$ 13 $\times$ 10$^8$ M$_\odot$ as the best estimate 
of the mass in the starburst region, with an upper limit of
20 $\times$ 10$^8$ M$_\odot$. Not only is this
value based on two independent and consistent determinations,
but also our extinction data indicate that the galaxy
is optically thin at the wavelengths and in the regions
critical for the mass determinations. Thus, the mass
should be measured reliability. 

\subsubsection{Mass in old stars}

To estimate the mass in old stars within the star forming 
region, we used the $K$-band fit to the 
surface brightness profile outside the ring of star formation 
(see Figure~7). The exponential
disk is assumed to account for the old stars present in the 
galaxy before the strong episode of star formation seen today.  
The disk fit was extrapolated to the center of the galaxy to 
estimate the contribution of old stars hidden by the
output of the starburst. No extinction correction
was applied, since the extinction to the disk is unknown
and in any case is likely to be less than the already modest
extinction (at $K$) to the nucleus. We estimated 1.2\,mJy of $K$-band
light is emitted by the old population within the central
2". Using the light-to-mass ratio derived
for spiral galaxy disks by Thronson \&
Greenhouse (1988) (corresponding to $M/L_V \sim 1.4$),
we find that the old stars account for
$M_{*,\,{\rm old}} \sim 12 \times 10^8\,{\rm M}_\odot$.

There is substantial evidence that the $M/L$ for spiral
disks is very similar from galaxy to galaxy and in
accordance with the values derived from Thronson \& Greenhouse
(e.g., Bottema 1993; 1999). In fact, the widely applied
Tully-Fisher relation probably depends on the
uniformity of $M/L$ in spiral disks (Aaronson, Mould,
\& Huchra 1979). Even if we assume that it is at the
extreme low side of the dispersion in the
Tully Fisher relation, the mass in the disk of NGC 1614
is unlikely to be more than 30\% lower than our estimate.
That is, assuming the lower limit $M/L_V\ge 1.1$,
a lower limit to the disk mass in the nuclear
region is 9 $\times 10^8\,{\rm M}_\odot$.
Our estimate ignores the mass of any bulge of old stars
hidden within the central starburst region, and hence
this lower limit is quite conservative.

Comparison of the mass in old stars
with the dynamical mass of 13 $\times$ 10$^8$ M$_\odot$ provides 
a difficult constraint. This problem cannot be relieved significantly
by enlarging the region under consideration. For example,
within a 3\arcsec\ diameter region, the flux from the
exponential disk model is 1.9\,mJy, corresponding to a
lower limit on the mass in old stars of 14.5 $\times$
$10^8\,{\rm M}_\odot$, which includes virtually all the
dynamical mass within this diameter also. 

\subsubsection{Mass of ISM from standard CO to H$_2$ conversion}

Use of the standard procedure for estimating the
mass in the interstellar medium from CO measurements
is strongly inconsistent with the mass budget
implied by the dynamics and the old stellar population; see Table~5.
Scoville et al. (1989) made observations of the nuclear
region in the $^{12}$CO J=1-0 line. These
results have been re-analyzed by Scoville et al. (1991)
and Bryant \& Scoville (1999); our discussion is based
primarily on the last reference. The nuclear CO source
is described as being unresolved with a diameter
$<$ 3\arcsec. Most of the material is probably
located in the ring of strong extinction revealed in
our $H - K$ images, which is 2\arcsec  ~in diameter.
Bryant \& Scoville (1999) use a "standard" conversion of CO emission to total
mass (H$_2$ + He) of molecular gas of N(H$_2$)/I$_{CO}$ = 2.25
$\times$ 10$^{20}$ cm$^{-2}$ (K km s$^{-1}$)$^{-1}$, 
corrected by a factor of 1.36 to allow for the mass in helium.
They deduce a nuclear molecular mass of 48 $\times$ 10$^8$ M$_\odot$,
nearly four times the total dynamical mass!  
Further details of the conversion (but with a slightly higher
standard factor) can be found in Sanders, Scoville, \& Soifer (1991).

The use of the standard conversion factor in intense
starbursts has come under question previously. Maloney \& Black (1988)
suggested that it should substantially overestimate the
molecular masses because of the relatively high densities
and temperatures of the molecular clouds in these regions.
Shier, Rieke, \& Rieke (1994) determined the 
dynamical mass with near infrared spectroscopy of
the 2.3$\mu$m CO bandhead in the nuclei of a
number of luminous starburst galaxies and
found that the masses of gas were overestimated
by the standard conversion by factors of $4 - 10$.
Solomon et al. (1997) and Downes \& Solomon (1998)
concluded the standard factor may be high
by a factor of five in starbursts, based on a comparison
of the dynamical mass and gas mass from far infrared measures.
Lisenfeld, Isaak, \& Hills (2000) used submillimeter measurements
of the emission by cold dust to argue that the CO-deduced mass
is typically $2-3$ times higher than would be deduced from the
thermal emission. Dunne et al. (2000) point out that such an increase is
one of the possible explanations for apparently large
values of M$_{H_2}$/M$_d$ in some strong merging/starburst galaxies. 
However, agreement has not been reached
as illustrated, for example, by the use of the standard
factor by Bryant \& Scoville (1996, 1999).

\subsubsection{Independent estimates of the molecular mass}

In NGC 1614, an independent test of the large molecular
mass deduced from the standard conversion can be based on the mm-wave CO
line profile. Casoli et al. (1991) have published a
profile in a 44\arcsec \  diameter beam that shows the full
width at 20\% intensity to be $\sim 300\,$km s$^{-1}$. The line
profile of Sanders et al. (1991) in a 55\arcsec\ beam has
a full width at 20\% intensity of 390 km s$^{-1}$. Scoville
et al. (1989) show that about 30\% of the emission
in these beams would have been from the central compact source. If
the mass of this region were $5 \times 10^9\,$M$_\odot$,
the line width would be $\sim 400\,$km s$^{-1}$ or greater for this
central 30\% component, which would be very difficult to reconcile
with the lack of strong, broad wings in the observed profile. 

We have made another estimate of the amount of
molecular gas, based on our discovery of an extinction
shadow due to a circumnuclear molecular ring.
We use the ratio of $^{18}$CO emission to
H$_2$ from the study of molecular clouds by Goldsmith,
Bergin, \& Lis (1997) and the relation between A$_V$ and
$^{18}$CO from Alves, Lada, \& Lada (1999) to relate
the extinction in this ring to its mass. From our data,
we estimate that the shadow of the molecular ring subtends
an annulus between radii of 0.5 and 1\arcsec \  with a typical
extinction level of A$_V$ $\sim 12\,$mag . Because the
extinction is only effective in the foreground of the
stellar population, we made the assumption that the
source is symmetric along the line of sight and therefore
doubled the mass estimates derived from the relations
cited above. Our value should probably be taken
as a lower limit: we would underestimate the extinction
if stars are mixed with the obscuring gas (or if some
stars lie in front of it, or the obscuration
is clumpy). Also, the reddening
at $H$ and $K$ in the thickest parts of the ring is strong
enough that optical depth effects might reduce the apparent
extinction. Nonetheless, our value of  $1-2$ $\times$ 10$^8$ M$_\odot$
is well aligned with the dynamical mass and, in fact, a
significantly higher value would create a serious problem
in explaining the nuclear properties given the other
constituents and their expected contributions to the mass.

An upper limit to the mass of interstellar material can be 
obtained from the discovery that this material is predominantly
distributed in a circumnuclear ring, as shown both by the 
Pa$\alpha$ morphology and the extinction shadow of a ring 
sector seen in the color maps. Although this ring is currently the site of vigorous star formation as the starburst propagates outward, 
it must have originally been a semi-stable configuration.  
The Toomre stability parameter for gas in a plane is 
\begin{equation}
Q = {\kappa \sigma \over \pi \Sigma G}
\end{equation}
where $\kappa$ is the epicyclic frequency,    $\sigma$
is the cloud velocity dispersion and $\Sigma$ is
the gas surface density. So that the gas remains in 
a disk and does not fragment into small clumps, $Q > 1$.
Thus, we can derive an upper limit for $\Sigma$:
\begin{equation}
\Sigma < 
470  M_\odot {\rm pc}^{-2}
\left({v_c \over 100 {\rm km/s}}\right)
\left({310 {\rm pc} \over r}\right)
\left({ \gamma \over 2   }\right)
\left({ \sigma \over 10 {\rm km/s}}\right)
\end{equation}
where $v_c$ is the  circular velocity and $\gamma = \kappa r/v_c$.
We have used a maximum value of $\gamma=2$ for a solid body rotation
curve. We assume a velocity dispersion of $\sigma = 20\,$km 
s$^{-1}$. This value was determined by extrapolating the observed
cloud size/line width relation (Falgarone, Puget, \&
P\'erault 1992) to a size of 100\,pc and then
doubling the width for good measure, to allow for possible
extreme conditions in a galaxy nuclear region.  
Taking $M_g = 2 \pi r dr \Sigma$ using
$dr = 150$ pc (based on the ring width in the extinction shadow),
we find a maximum mass of $M_g < 4 \times 10^8 M_\odot$.
If there is a high dispersion component of material, the actual
gas mass could be somewhat higher.

Another lower limit of M$_g$ $>$ 2 $\times$ 10$^8$ M$_\odot$
can be computed by assuming the CO emission is optically thin
(Bryant \& Scoville 1996, equation A4). 

\subsubsection{Conclusions about the ISM in the nucleus of NGC 1614}

Combining these arguments, the molecular mass in the nucleus is most likely
$\sim$ 3 $\times$ 10$^8$ M$_\odot$, about 25\% of the dynamical
mass in this region. This mass is about 15 times lower than the 
$4.8 \times 10^9 M_\odot$ in molecular mass 
deduced from CO and the standard conversion by Bryant and Scoville (1999).
Table~5 illustrates how tight the mass budget is. It
seems unlikely that the molecular gas can be
more than 1/10 the amount estimated from the
standard conversion even if the dynamical mass is set
to the upper limit of 20 $\times$ 10$^8$ M$_\odot$ and 
the mass in old stars to its lower limit of 9 $\times$ 10$^8$ M$_\odot$. In that
case, it might be necessary to incorporate a significant
change in the starburst initial mass function (see
Section 3.5.2). Our results are in agreement with those of Shier 
et al. (1994), who estimated that the amount
of gas in the nucleus of NGC 1614 was a factor of at least
nine below the value from the standard conversion.

To accompany their estimate of a huge mass of molecular gas,
Scoville et al. (1989) inferred the extinction in the nuclear
region of NGC~1614 to be $A_V >95\,$mag.
They argued that much of the "action" in this region would
be hidden from optical and even infrared observers unless
there happened to be a particularly favorable line
of sight. This picture of LIRGs and ULIRGs has gained
wide credence, but the observations presented here
are a serious caution. Not only do we find far less molecular
mass, but more significantly for the "hidden action" model,
we {\it do not} find anything approaching the high extinction
deduced from the CO observations. An independent indication
that the extinction is not hiding any significant action
is the close correspondence between the star formation
rates deduced from Pa$\alpha$ and far infrared luminosities,
as discussed in Section 3.2.
     
\subsection{The star formation properties}

For our analysis of the star formation properties of NGC~1614,
we will use the Rieke et al. (1993) evolutionary synthesis
starburst models because they include
a careful calibration of the CO stellar absorption band
strengths against observational data. We have considered
Gaussian bursts of 5\,Myr duration (FWHM) with the peak of
star formation at 5\,Myr after the beginning time for the burst.
NGC~1614 is a well known Wolf-Rayet (WR) galaxy (Vacca \& Conti
1992; Schaerer, Contini, \& Pindao 1999). The presence of
WR features in star forming galaxies can only be explained with short
bursts of star formation with durations of the order of a few million
years (Leitherer \& Heckman 1995), as used in our fit to
the observed properties of the galaxy.

We used a truncated Salpeter
IMF\footnote{$\phi(m){\rm d}m \propto m^{-2.35}{\rm d}m$
from 1 to $80\,{\rm M}_\odot$, and $\phi(m){\rm d}m \propto
m^{-1}{\rm d}m$ from 0.1 to $1\,{\rm M}_\odot$.}
which is virtually identical to
IMF8 which Rieke et al. (1993) decided gave the best fit
to the properties of M82, and has since
been shown to fit other starburst galaxies well (Engelbracht et al. 1996,
1998). In Table~6 we summarize the properties of the central starburst 
of NGC~1614 that will be fitted with the evolutionary synthesis models.

\begin{deluxetable}{llll}
\tablewidth{13cm}
\tablefontsize{\small}
\tablecaption{Starburst model constraints for the star forming nuclear ring.}
\tablehead{\colhead{Property} &  \colhead{Value} & \colhead{Origin}&
\colhead{Reference}}
\startdata
Dynamical Mass & $1.3 \times 10^9\,{\rm M}_\odot$  & IR spectroscopy &
this work \\
Infrared Luminosity & $3\times 10^{11}\,{\rm L}_\odot$ & IRAS &
Goldader et al. (1997) \\
Ionizing photons (log)   & 54.5, $54.6\,{\rm s}^{-1}$
& [Ne\,{\sc ii}], Pa$\alpha$ & Roche et al. (1991), this work\\
$K$ absolute magnitude & $-23.6\,$mag & \nodata &Engelbracht (1997)\\
SNr           & $0.3\,{\rm yr}^{-1}$ & [Fe\,{\sc ii}] &
this work\\
CO index & 0.23 & IR spectroscopy & this work\\
\enddata
\tablecomments{The values of $N_{\rm Ly}$ (from
Pa$\alpha$), $K$-band absolute magnitude and
SNr are corrected using an extinction of $A_K = 0.5\,$mag.}
\end{deluxetable}

The infrared luminosity will be
taken as a lower limit to the bolometric luminosity of the galaxy.
This assumption makes two errors in opposite
direction. From the aperture measurements discussed
in Section 3.2, up to about one third of the
infrared luminosity of the galaxy may be produced
outside the nuclear region. On the other hand,
a significant amount of optical/UV energy must escape from
the starburst before it is absorbed and re-emitted in
the infrared, particularly given the patchy
and relatively low level of extinction.
The number of ionizing photons is from both the
[Ne\,{\sc ii}]$12.8\,\mu$m line flux of Roche et al. (1991),
and the {\it HST}/NICMOS Pa$\alpha$ flux,
the latter corrected for an average extinction of
$A_K=0.5\,$mag and assuming case B recombination. Our estimate of
the number of ionizing photons is in good agreement with the
value from Puxley \& Brand (1994) derived from measurements
that included Br$\alpha$, analyzed through detailed modeling
of the extinction. It is also consistent with the upper
limit established from the 3$\sigma$ non-detection of the 
radio H$\alpha$92 hydrogen recombination line 
(Phookun, Anantharamaiah, \& Goss 1998). To
allow for the possibility of absorption of ionizing
photons by interstellar dust, in the models we take
the derived UV flux as a lower limit. 
The [Fe\,{\sc ii}] line ratios are similar to those
detected in NGC~253 (Engelbracht et al. 1998) and
are typical of those in supernova remnants. 
We estimated the supernova rate (SNr) from the
[Fe\,{\sc ii}]$1.644\,\mu$m luminosity (corrected for 
extinction) from this work and a new calibration between
these two quantities (Alonso-Herrero et al. 2000b, in preparation).

\begin{figure*}
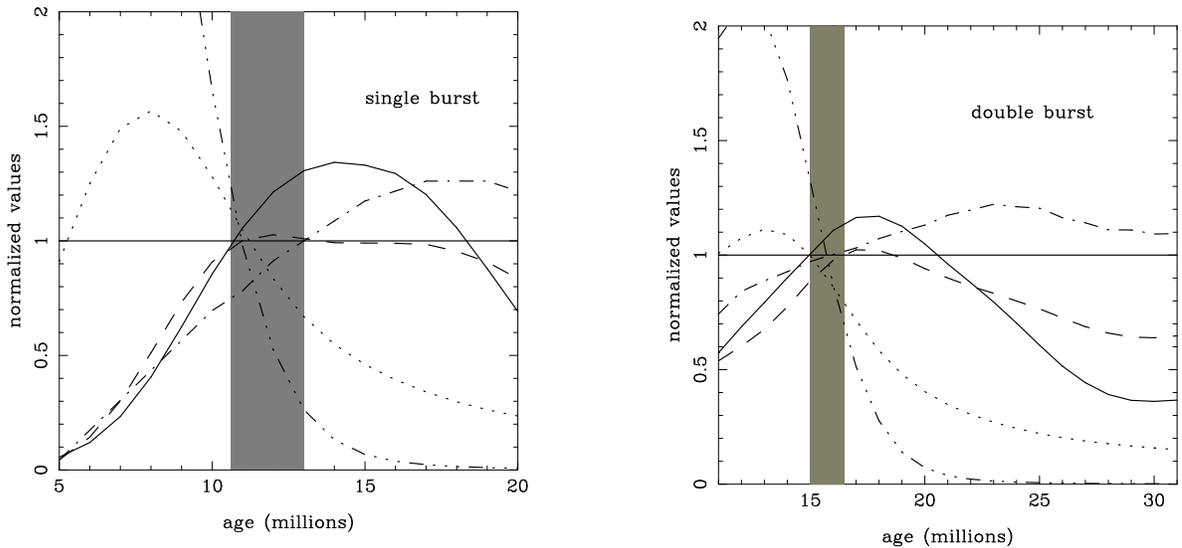

\figurenum{8}
\plotfiddle{figure8a.ps}{425pt}{0}{60}{60}{-250}{170}
\plotfiddle{figure8b.ps}{425pt}{0}{60}{60}{0}{600}
\vspace{-22cm}
\caption{{\it Left panel}. Best fit to 
the properties of the central
starburst of NGC~1614 for the short-duration Gaussian burst ${\rm FWHM} =
5\times 10^6\,$yr. {\it Right panel}. Best fit using a
double Gaussian burst (${\rm FWHM} = 5\times 10^6\,$yr). The elapsed
between the two bursts is $5\times 10^6\,$yr.
In both Figures~8a and 8b, the solid line is the $K$-band
luminosity, the dashed line is the SNr, the dot-dash line
is the CO index, the dotted line is the bolometric
luminosity and the dot-dot-dot-dash line is the
number of ionizing photons. The age in the x-axis is the
age from the beginning of the star formation episode. The
peak of star formation occurs after 5\,Myr. The shaded area
shows the limiting starburst ages for a given fit.}
\end{figure*}

Our best fit to the properties of NGC~1614 with a single burst of
star formation is presented in Figure~8 (left panel). In this diagram
the output of the model is normalized to the properties of NGC~1614 so
the target quantity is unity. The model uses a
mass in newly formed stars of $5.6 \times 10^8\,{\rm M}_\odot$
($\simeq 0.4 \times M_{\rm dyn}$). The derived age is between
5\,Myr and 8\,Myr after the peak of star formation,
or $7-11\,$Myr measured from the rising
half power of the assumed Gaussian star formation rate. This 
age is consistent with our failure to detect the 1.7$\mu$m
He line, which suggests (although it is not a strong
result given the signal to noise) that there are relatively few
stars hotter than 40,000\,K (Vanzi et al. 1996). All the observables except
the CO index converge at the young age. The observed CO index appears
to be too deep for such a young population.

The strong CO of this galaxy is more typical of a stellar population
ten million years old or older. The discrepancy with
the age of 5 $-$ 8 Myr for the other starburst indicators
suggests that this galaxy has experienced a more extended
period of star formation than in our single burst. 
The simplest example would be a double burst. In
Figure~8 (right panel) we show the best fit
model obtained with two Gaussian bursts (both with FWHM = 5\,Myr)
separated by 5\,Myr. Each burst uses 20\% of the dynamical mass. As can
be seen from this figure, the fit is excellent. The derived age is
$10-11.5\times 10^6\,$yr after the peak of star formation of the first
burst. A variety of other time dependencies of star formation
are also likely to fit the data, but the minimum duration of the
episode and the minimum mass required are both
well constrained by the double burst model.

This starburst model coincides closely with
the morphology of the galaxy. We find that the strong
CO is concentrated in the nucleus (Figure~4, upper panel), and the
modeling associates this feature with the older part
of the starburst. Most of the ionizing flux is
produced in the younger part of the starburst, and
is physically located in the circumnuclear ring (Figure~6). 
From the starburst modeling and the diameter of the circumnuclear
ring of H\,{\sc ii} regions, we can roughly estimate that the speed at
which the starburst is propagating outward is $\sim 60\,$km s$^{-1}$.

\subsection{The Initial Mass Function}

\subsubsection{Previous estimates of the IMF in starbursts}

Rieke et al. (1980) found that evolutionary starburst
models using the then-current form for
the local IMF (Miller \& Scalo 1979) had difficulty fitting
the properties of M82 within the available dynamical mass.
They suggested that the IMF in the starburst might be
biased toward massive stars compared with the proposed local form.
Since the low mass stars contribute most of the total mass
in the IMF, but have no observable properties in a starburst,
reducing their numbers is a plausible way to reconcile the masses.
This possibility has been discussed at great length since.
Rieke et al. (1993) reconsidered M82, experimenting with
a broad variety of forms for the IMF and histories of star formation.
They confirmed that the currently proposed forms for the local IMF (e.g.,
Scalo 1986) were inadequate and proposed their "IMF8" as
a more satisfactory solution, IMF8 provides $4-5$  times 
more stars in the $10-30\,{\rm M}_\odot$ range than does Scalo (1986)
but drops toward higher masses to avoid overproduction of
oxygen in very massive stars (Rieke et al. 1993).

In the last few years, modeling of the ultraviolet spectra
of starburst galaxies has led to the conclusion that their
IMFs have roughly the Salpeter slope at high masses
(e.g., Gonz\'alez-Delgado et al. 1999).
Satyapal et al. (1997) modeled M82 with less complete
evolutionary models than used by Rieke et al. (1993).
They claimed that they could account for
the $K$-band luminosity with a Salpeter IMF, avoiding the necessity
to suppress formation of low mass stars compared
with the local IMF. A problem with their
approach is that they did not demonstrate the
simultaneous fitting of all the starburst
constraints used by Rieke et al. (1993); it is
relatively easy to fit a subset of constraints if
the rest are ignored. Furthermore, because they
assumed a foreground screen extinction model,
the $K$-band luminosity they fitted is in fact a
lower limit to the true luminosity, making
their fits less demanding. Nonetheless, their success
with a modified Salpeter IMF confirms independently
the derivation of IMF8 to fit M82 by Rieke et al. (1993),
as we show below.

\begin{deluxetable}{lcc}
\tablewidth{8cm}
\tablefontsize{\small}
\tablecaption{Percentage of Mass in High Mass Stars.}
\tablehead{\colhead{IMF} &  \colhead{10 $-$ 30 M$_\odot$} &
\colhead{30 $-$ 100 M$_\odot$}\\}
\startdata
IMF8              & 14\% & 5\% \\
Modified Salpeter  & 13\% & 9\% \\
Original Salpeter  & 7\%  & 5\% \\
Miller \& Scalo 1979  & 6.5\%& 3\%  \\
Scalo 1986         & 3\%  & 1\%  \\
Basu \& Rana 1992  & 3\%  & 3.5\%  \\
\enddata
\end{deluxetable}

The critical aspect of a starburst IMF constrained
by luminosities and dynamical mass is the proportion of
mass it places in the $10-30\,{\rm M}_\odot$ range
that dominates the observable aspects of the starburst.
Table~7 summarizes this proportion and that of more
massive stars for the various forms of the IMF
that have been used to fit starbursts. It shows
the similarity of IMF8 and the modified Salpeter IMF,
which are equivalent in fitting the primary
starburst characteristics. However, IMF8
is to be preferred because its factor of two
lower portion of very massive stars results
in a proportionate reduction in the production
of oxygen. Keeping the abundance of oxygen within
the observational limits is a difficult constraint in
starburst models (Rieke et al. 1993; Wang \& Silk 1993). The other
forms for the IMF produce significantly fewer
10 $-$ 30 M$_\odot$ stars. To first order, the
dynamical masses required with them are just
in the ratio of the proportions of these
stars. Thus, if M82 can be fitted with 2.5 $\times$
$10^8\,{\rm M}_\odot$ in a modified Salpeter IMF, it
would require about $5 \times 10^8\,$M$_\odot$
with a Miller-Scalo IMF, about 70\% of the dynamical
mass available and unlikely to be an acceptable fit
once allowance is made for the mass of other constituents
of the nuclear region. For NGC~1614, the Miller-Scalo
IMF would take all the dynamical mass and the Scalo (1986)
and Basu \& Rana (1992) ones would require about twice the
dynamical mass, all of which is clearly unacceptable.

Virtually all studies find the local IMF
to fall more steeply toward high
masses than the Salpeter form.
Even Salpeter (1955) continued the simple power law to high
masses only for simplicity, but at the same time
called attention to an apparent high mass steepening:
"It is not yet clear whether the steeper
drop for masses larger than 10 M$_\odot$
is a real effect, since in this region masses and bolometric 
corrections are not known very accurately and the
number of such stars near the galactic plane is small." 
As with the rest of his work this
statement has proven prescient. However, a Salpeter IMF
has been shown to be an approximate fit in many
situations outside the local neighborhood (e.g., Scalo 1998).
It now appears that a better description of the M82 issue
might be that the local IMF (or at least
our estimates of it) may differ from the IMF we find
in most regions of active star formation, including
in starbursts like M82.

\subsubsection{NGC~1614: a more extreme IMF?}

We have emphasized that NGC~1614 is particularly well suited
to studying starburst properties, given the relatively
modest near infrared extinction and our favorable
viewing of it at modest inclination. It has
a well determined dynamical mass of 13 $\times$ 10$^8$
M$_\odot$ in the starburst region, as discussed in Section 3.2.
However, the constituents in this region appear to sum
to a significantly greater mass, $>$ 9 $\times$ 10$^8$ M$_\odot$
for the old stars plus $\sim$ 3 $\times$ 10$^8$ M$_\odot$ for the
molecular gas plus 5.5 $\times$ 10$^8$ M$_\odot$ for the
starburst using either a Salpeter IMF or IMF8. The total
exceeds the dynamical mass by at least 35\%. Unless
there is substantial contamination of the disk
outside the starburst by younger stars, these numbers suggest
that the mass in the starburst may be overestimated. A
further indication in the same direction is that the
starburst plus remaining molecular gas in our nominal
estimates account for 65\% of the dynamical mass. 
Theoretical modeling suggests that instabilities
will set in and a starburst will appear once the mass
of gas reaches $\sim$ 20\% of the total in a galaxy nucleus
(Wada \& Habe 1992; Bekki 1995). These estimates
are somewhat uncertain and may not apply strictly to the
conditions in NGC~1614. However, they fall so far below the deduced
mass of new stars and residual gas to suggest that
the IMF in the NGC~1614 starburst is weighted
toward massive stars, even compared
with IMF8 or a modified Salpeter function. This
behavior should be tested in studies of other extreme
starbursts. 

\subsection{The star formation efficiency}

The mechanism needed to account for the
strong star formation in luminous mergers is a matter of
debate. The two main competing models are:
(1) cloud-cloud collisions which predict
a large population of high mass stars to give rise to the
high infrared luminosities (Scoville, Sanders, \& Clemens 1986)
(2) gravitational instability in nuclear gas
disks, as favored by Taniguchi \&
Ohyama (1998). The former model would predict a Schmidt
law\footnote{The star formation rate density is
a gas density power law, $\Sigma_{\rm SFR} \propto \Sigma_{\rm gas}^N$,
Schmidt (1959)} with
index $N=2$, whereas the latter model predicts indices $N=1-1.5$ as
found for isolated galaxies and strong starbursts (Kennicutt 1998;
Taniguchi \& Ohyama 1998).

From the Pa$\alpha$ luminosity of NGC~1614 (corrected
for extinction) and using the relation between SFR and H$\alpha$
luminosity (Kennicutt 1998) we measured a ${\rm SFR} = 36\,{\rm M}_\odot
\,{\rm yr}^{-1}$ in the ring and a SFR surface density of
$\Sigma_{\rm SFR} = 120 \pm 50\,{\rm M}_\odot\,
{\rm yr}^{-1}\,{\rm kpc}^{-2}$, similar to,
although  slightly higher than, the earlier estimate by Kennicutt (1998)
from the FIR luminosity and somewhat different parameters for
the size of the emitting region
($\Sigma_{\rm SFR} = 65\,{\rm M}_\odot\,{\rm yr}^{-1}\,{\rm kpc}^{-2}$).
The value of $\Sigma_{\rm SFR} =
120\,{\rm M}_\odot\,{\rm yr}^{-1}\,{\rm kpc}^{-2}$ that we found
for NGC~1614 is of the same magnitude as that of the giant H\,{\sc ii}
region 30 Doradus (Kennicutt 1998), and the global
star formation efficiency for this galaxy approaches 100\% over a
gas consumption period of $10^8\,{\rm yr}$.
Based on the standard conversion, Kennicutt (1998) derived
$N = 1.4 \pm 0.15$. If the mass in molecular
gas is only $\sim$ 3 $\times$ 10$^8$ M$_\odot$,
with a conversion of CO to H$_2$ mass a factor of
$\sim$ 15 lower than the standard value,
one would derive $N = 2$ or more, as discussed by Kennicutt (1998)
(with less gas, higher efficiency is needed).
Thus, our results appear to favor the cloud collision
explanation for the high rate of star formation.

\subsection{Comparison with Simulations}

Mihos \& Hernquist (1994) explored via numerical
simulations the merger of a large and a smaller galaxy, similar
to our proposal for NGC~1614. From their simulation they predicted that
in response to the tidal perturbation of the infalling satellite, the
disk galaxy develops strong spiral arms. This non-axisymmetric perturbation
causes large quantities of material
to be funneled into the central regions, and
as a result a strong burst of star formation is triggered lasting for
up to 60\,Myr. The predicted
central starburst is very compact ($\simeq 350\,$pc
in radius) and contains some 85\% of the total mass in newly formed
stars.

Hernquist \& Mihos (1995) carried out further simulations.
They showed that the degree of central concentration of
gas decreases markedly with the presence of a massive
bulge. The relatively low dynamical mass in the central
620 pc of NGC~1614 indicates that it had little bulge prior
to the collision, so it would be ideal for a strong central
concentration of gas. Hernquist \& Mihos (1995) state that
the resolution of their calculations is inadequate to
determine how the gas behaves once it has sunk into the
central few hundred pc. However, from the behavior of
NGC~1614, it appears that it must be further concentrated
until instability is reached in the bottom of the potential
well of the galaxy, leading to the initial stages of
massive star formation.

As the starburst ages, supernovae begin to explode and
winds develop that expand outward, compressing the
remaining molecular gas surrounding the nucleus (e.g.,
Tenorio-Tagle \& Mu\~noz-Tu\~n\'on 1997; Taniguchi, Trentham, \& Shioya 1998).
Gravitationally unstable shocked layers of interstellar gas
may be produced in cloud cloud collisions and by the impact
of these winds on incoming clouds. The
resulting star formation appears to
favor massive stars (Whitworth et al. 1994),
possibly accounting for the suggestion of a
high abundance of such stars in extreme starbursts.
The evidence we find that the starburst in NGC~1614
has expanded from the nucleus to the ring of H\,{\sc ii} regions
and into the surrounding molecular ring is in agreement
with this picture.
 
\section{SUMMARY}

We have used {\it HST}/NICMOS and WFPC2 imaging, ground-based
near infrared spectra, and information from the literature
to study the luminous infrared galaxy NGC~1614.
The modest extinction to its starbursting nucleus,
combined with our favorable viewing angle, make this
galaxy ideal for probing the properties of its powerful starburst.
We show that

\itemize

\item Star formation has propagated outward from the nucleus to
a $\sim 300\,$pc radius circumnuclear ring of massive H\,{\sc ii} regions.

\item These circumnuclear H\,{\sc ii} regions are an order of magnitude
more luminous than 30 Doradus, and H\,{\sc ii} regions in
the spiral arms are similar in luminosity to 30 Dor. These
regions are probably younger versions of the luminous
stellar clusters also around the nucleus and in the spiral arms. The
presence of so many extremely luminous H\,{\sc ii} regions and
stellar clusters is similar to the behavior of other luminous starbursts.

\item The ring of H\,{\sc ii} regions is surrounded by molecular gas.
The mass in molecular gas appears to be $\sim 1/15$ the
prediction from the standard $^{12}$CO/H$_2$ ratio, with
an upper limit of 1/10 the standard ratio. The
large levels of extinction predicted previously from the 
standard ratio are absent in NGC~1614.

\item The mass budget for the starburst is very tight, suggesting
that the proportion of massive stars is larger than predicted
by a Salpeter IMF modified to turn over near 1 M$_\odot$,
or by the virtually identical IMF8 found to
fit the starburst in M82 (Rieke et al. 1993).
Until now, such IMFs have been adequate to fit the properties
of all known starbursts.

\item The high efficiency needed to convert
the available molecular gas density
into the observed density of star formation suggests that
the starburst is propagating by cloud-cloud interactions
and winds.

\item The behavior of NGC~1614 is consistent with simulations
for the interaction of a modest sized galaxy with a massive,
late type spiral. 

\enditemize

\section*{Acknowledgments}

We thank J. Hinz for a helpful discussion and the
anonymous referee for a thorough critique of
the first version of this paper. This work was supported by the National
Aeronautics and Space Administration through grant NAG 5-3042
and by the National Science Foundation Grant AST-9529190.


\begin{thebibliography}{}

\bibitem[Aaronson, Mould, \& Huchra \
1979]{aar} Aaronson, M., Mould, J., \& Huchra, J. 1979, ApJ, 229, 1

\bibitem[Alonso-Herrero et al. \
2000a]{almu} Alonso-Herrero, A., Rieke, M. J., Rieke, G. H., \&
Scoville, N. Z. 2000a, ApJ, 532, 845

\bibitem[Alonso-Herrero et al. \
2000b]{almub} Alonso-Herrero, A., Rieke, M. J., Rieke, G. H., \&
Kelly, D.  2000b, in preparation

\bibitem[Alves, Lada, \& Lada \
1999]{all} Alves, J., Lada, C. J., \& Lada, E. A. 1999, ApJ, 515, 265

\bibitem[Barnes \& Hernquist \
1996]{bar} Barnes, J. E., \& Hernquist, L. 1996, ApJ, 471, 115

\bibitem[Basu \& Rana \
1992]{bas92} Basu, S., \& Rana, N. C. 1992, ApJ, 393, 373

\bibitem[Bekki \
1995]{bek} Bekki, K. 1995, MNRAS, 276, 9

\bibitem[Black \& van Dishoeck,  \
1987]{bla} Black, J. H. \& van Dishoeck, E. F. 1987, ApJ, 322, 412 

\bibitem[Bottema \
1993]{bot93} Bottema, R. 1993, A\&A, 275, 16

\bibitem[Bottema \
1999]{bot99} Bottema, R. 1999, A\&A, 348, 77

\bibitem[Bryant \& Scoville \
1996]{bry96} Bryant, P. M., \& Scoville, N. Z. 1996, ApJ, 457, 678
 
\bibitem[Bryant \& Scoville \
1999]{bry99} Bryant, P. M., \& Scoville, N. Z. 1999, AJ, 117, 2632

\bibitem[Carico et al. \
1988]{car} Carico, D. P., Sanders, D. B., Soifer, B. T., Elias, J. H.,
Matthews, K., \& Neugebauer, G. 1988, AJ, 95, 356

\bibitem[Casoli et al. \
1991]{cas} Casoli, F., Dupraz, C., Combes, F., \& Kazer, I.
1991, A\&A, 251, 1

\bibitem[Downes \& Solomon \
1988]{down} Downes, D., \& Solomon, P. D. 1998, ApJ, 507, 615

\bibitem[Dunne et al.\
2000]{dun} Dunne, L., Eales, S., Edmunds, M., Ivison, R., Alexander, P., 
\& Clements, D. L. 2000, MNRAS, 315, 115

\bibitem[Engelbracht et al. \
1996]{engel96} Engelbracht, C. W., Rieke, M. J., Rieke, G. H., \&
Latter, W. B. 1996, ApJ, 467, 227

\bibitem[Engelbracht et al. \
1998]{engel98} Engelbracht, C. W., Rieke, M. J., Rieke, G. H., Kelly,
D. M. \& Achtermann, J. M. 1998, ApJ, 505, 639

\bibitem[Falgarone et al. \
1992]{fal92} Falgarone, E., Puget, J.-L., \& P\'erault, 
M. 1992, A\&A, 257, 715

\bibitem[Goldader et al. \
1997]{gol} Goldader, J. D., Joseph, R. D., Doyon, R., \& Sanders, D.
B. 1997, ApJ, 474, 104

\bibitem[Goldsmith et al. \
1997]{gbl} Goldsmith, P. F., Bergin, E. A., \& Lis, D. C. 1997,
ApJ, 491, 615

\bibitem[Gonz\'alez-Delgado et al. \
1999]{gon} Gonz\'alez-Delgado, R. M., Garc\'{\i}a-Vargas, M. L.,
Goldader, J., Leitherer, C., \& Pasquali, A. 1999, ApJ, 513, 707

\bibitem[Griffith et al. \
1995]{grif} Griffith, M. R., Wright, A. E., Burke, B. F., \&
Ekers, R. D. 1995, ApJS, 97, 347

\bibitem[Hernquist \& Mihos \
1995]{her} Hernquist, L. \& Mihos, J. C. 1995, ApJ, 448, 41
 
\bibitem[Ho et al. \
1990]{ho} Ho, P. T. P., Beck, S. C., \& Turner, J. L. 1990, 
ApJ, 349, 57

\bibitem[Kennicutt, Edgar, \& Hodge \
1989]{keh} Kennicutt, R. C. Jr., Edgar, B. K., \&
Hodge, P. W. 1989, ApJ, 337, 761

\bibitem[Kennicutt \
1998]{ken} Kennicutt, R. C. Jr. 1998, ApJ, 498, 541

\bibitem[Keto et al. \
1992]{keto} Keto, E., Ball, R., Arens, J., Jernigan, G., \&
Meixner, M. 1992, ApJ, 389, 223

\bibitem[Lebofsky \& Rieke \
1979]{leb} Lebofsky, M. J., \& Rieke, G. H. 1979, ApJ, 229, 111

\bibitem[Leitherer \& Heckman \
1995]{lei} Leitherer, C., \& Heckman, T. M. 1995, ApJS, 96, 9

\bibitem[Lisenfeld et al. \
2000] {lis} Lisenfeld, U., Isaak, K. G., \& Hills, R.
2000, MNRAS, 313, 433

\bibitem[Lutz et al. \
1999]{lutz} Lutz, D., Spoon, H. W. W., Rigopoulou, D.,
Moorwood, A. F. M., \& Genzel, R. 1999, ApJ, 505, L103

\bibitem[Maloney \& Black \
1988]{mal} Maloney, P., \& Black, J. H. 1988, ApJ, 325, 389

\bibitem[Mazzarella \& Boroson \
1883] {mazz} Mazzarella, J. M., \& Boroson, T. A. 1993,
ApJS, 85, 27

\bibitem[Mihos \& Hernquist \
1994]{mihos} Mihos, J. C., \& Hernquist, L. 1994, ApJ, 425, L13

\bibitem[Miles et al. \
1996]{miles} Miles, J. W., Houck, J. R., Hayward, T. L., \&
Ashby, M. L. N. 1996, ApJ, 465, 191

\bibitem[Miller \& Scalo \
1979]{mil79} Miller, G. E., \& Scalo, J. M. 1979,
ApJS, 54, 513

\bibitem[Neff et al. \
1990]{neff} Neff, S. G., Hutchings, J. B., Stanford, S. A., \&
Unger, S. W. 1990, AJ, 99, 1088

\bibitem[Phookun et al. \
1998]{ph} Phookun, B., Anantharamaiah, K. R., \& Goss, W. M. 
1998, MNRAS, 295, 156

\bibitem[Puxley \& Brand 1994 \
1994]{pux94} Puxley, P. J., \& Brand, P. W. J. L. 1994, MNRAS, 266, 431

\bibitem[Puxley \& Brand  \
1999]{pux99} Puxley, P. J., \& Brand, P. W. J. L. 1999, ApJ, 514, 675
           
\bibitem[Rieke \& Low \
1972]{rie72} Rieke, G. H., \& Low, F. J. 1972, ApJ, 176, L95

\bibitem[Rieke et al. 1980 \
1980]{rie80} Rieke, G. H., Lebofsky, M. J., Thompson, R. I.,
Low, F. J., \& Tokunaga, A. T. 1980, ApJ, 238, 24

\bibitem[Rieke \& Lebofsky \
1985]{rie85} Rieke, G. H., \& Lebofsky, M. J. 1985, ApJ, 288, 618

\bibitem[Rieke et al. 1993 \
1993]{rie93} Rieke, G. H., Loken, K., Rieke, M. J., \&
Tamblyn, P. 1993, ApJ, 412, 99

\bibitem[Ridgway, Wynn-Williams, \& Becklin \
1994]{rid94} Ridgway, S. E., Wynn-Williams, C. G., \&
Becklin, E. E. 1994, ApJ, 428, 609

\bibitem[Roche et al. \
1991]{roche} Roche, P. F., Aitken, D. K., Smith, C. H., \&
Ward, M. J. 1991, MNRAS, 248, 606

\bibitem[Rozas et al. \
1996]{rozas} Rozas, M., Beckman, J. E., \& Knapen, J. H. 1996,
A\&A, 307, 735

\bibitem[Salpeter \
1955]{sal} Salpeter, E. E. 1955, ApJ, 121, 161

\bibitem[Sanders et al. \
1991]{san91} Sanders, D. B., Scoville, N. Z., \& Soifer, B. T.
1991, ApJ, 370, 158

\bibitem[Sanders et al. \
1988]{san88} Sanders, D. B., Soifer, B. T., Elias, H. J., Modore, B. F.,
Mathews, K., Neugebauer, G., \& Scoville, N. Z. 1988, ApJ, 325, 74

\bibitem[Sanders \& Mirabel \
1996]{samir} Sanders, D. B., \& Mirabel, I. F. 1996, ARA\&A, 34, 749

\bibitem[Satyapal et al. \
1997]{sat} Satyapal, S., Watson, D. M., Pipher, J. L.,
Forrest, W. J.,  Greenhouse, M. A., Smith, H. A.,
Fischer, J., \& Woodward, C. E. 1997, ApJ, 483, 148

\bibitem[Scalo \
1986]{sca86} Scalo, J. M. 1986, Fund. Cosmic Phys., 11, 1

\bibitem[Scalo \
1998]{sca98} Scalo, J. 1998, in The Stellar Initial Mass Function 
(38th Herstmonceux Conference) edited by Gary Gilmore 
and Debbie Howell. ASP Conference Series, Vol. 142, 1998, p.201

\bibitem[Schaerer et al. \
1999]{sch} Schaerer, D., Contini, T. \& Pindao, M. 1999, A\&AS, 136, 358

\bibitem[Schmidt \
1959]{schm} Schmidt, M. 1959, ApJ, 129, 243

\bibitem[Scoville et al. \
1999]{sco99} Scoville, N. Z., Evans, A. S., Thompson, R., Rieke,
M., Hines, D., Low, F. J., Dinshaw, N., Surace, J. A., \&
Armus, L. 2000, AJ, 119, 991

\bibitem[Scoville et al. \
1991]{sco91} Scoville, N. Z. Sargent, A. I., Sanders, D. B., \&
Soifer, B. T. 1991, ApJ, 366, L5

\bibitem[Scoville et al. \
1989]{sco89} Scoville, N. Z., Sanders, D. B., Sargent, A. I.,
Soifer, B. T., \& Tinney, C. G. 1989, ApJ, 345, L25

\bibitem[Scoville et al. \
1986]{sco86} Scoville, N. Z., Sanders, D. B., \&
Clemens, D. P. 1986, ApJ, 310, L77

\bibitem[Shier et al. \
1994]{shier94} Shier, L. M., Rieke, M. J., \& Rieke, G. H. 1994,
ApJ, 433, L9

\bibitem[Shier et al. \
1996]{shier} Shier, L. M., Rieke, M. J., \& Rieke, G. H. 1996,
ApJ, 470, 222

\bibitem[Solomon et al. \
1997]{solo} Solomon, P. M., Downes, D., Radford, S. J., E., 
\& Barrett, J. W. 1997, ApJ, 478, 144

\bibitem[Sternberg \& Dalgarno  \
1989]{stern} Sternberg, A., \& Dalgarno, A. 1989, ApJ, 338, 197

\bibitem[Taniguchi \& Ohyama \
1998]{tani} Taniguchi, Y., \& Ohyama, Y. 1998, ApJ, 509, L89

\bibitem[Taniguchi, Trentham, \& Shioya \
1998]{tantren} Taniguchi, Y., Trentham, N. \& Shioya, Y.
1998, ApJL, 504, L79

\bibitem[Tenorio-Tagle \& Mu\~noz-Tu\~n\'on \
1997]{ten} Tenorio-Tagle, G., \& Mu\~noz-Tu\~n\'on, C. 1997, ApJ, 478, 134

\bibitem[Thronson \& Greenhouse \
1988]{thr} Thronson, H. A. Jr., \& Greenhouse, M. A. 1988, 
ApJ, 327, 671

\bibitem[Vacca \& Conti \
1992]{vacca} Vacca, W. D., \& Conti, P. S. 1992, ApJ, 401, 543

\bibitem[Vanzi et al. \
1996]{van} Vanzi, L., Rieke, G. H., Martin, C. L., \& Shields, J. C.
1996, ApJ, 366, 150

\bibitem[Wada \& Habe \
1992]{wad} Wada, K., \& Habe, A. 1992, MNRAS, 258, 82

\bibitem[Wang \& Silk \
1993]{wang}Wang, B., \& Silk, J. 1993, ApJ, 406, 580

\bibitem[Whitworth et al. \
1994]{whit} Whitworth, A. P., Bhattal, A. S.,
Chapman, S. J., Disney, M. J., \& Turner, J. A. 1994,
MNRAS, 268, 291
            
\bibitem[Williams et al. \
1993]{wi}Williams, D. M., Thompson, C. L., Rieke, G. H., \&
Montgomery, E. F. 1993, S.P.I.E., 1946, 482

\bibitem[Witt et al. 1992 \
1992]{witt} Witt, A. N., Thronson, H. A. Jr., \& Capuano, J. M. Jr.
1992, ApJ, 393, 611

\end{thebibliography}
\end{document}